\documentclass[fontsize=11pt,headings=small,headinclude=true,footinclude=true, 
 headings=optiontohead,toc=listof,DIV=22,listof=leveldown, headsepline , letterpaper]{scrartcl} 
\usepackage[utf8]{inputenc}
\usepackage[T1]{fontenc}
\usepackage[ngerman, english]{babel}

\newcommand{\dptitle}{Modeling the out-of-equilibrium dynamics \\ of bounded rationality and economic constraints}
\newcommand{\dptitleclean}{Modeling the out-of-equilibrium dynamics of bounded rationality and economic constraints}

\newcommand{\dpautoren}{Oliver Richters \href{https://orcid.org/0000-0001-8253-4716}{\includegraphics[width=1em]{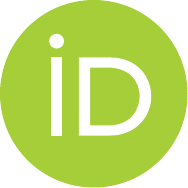}}}
\newcommand{\dpautorenclean}{Oliver Richters}

\newcommand{\dpaffiliation}{\small International Economics, Department of Economics, Carl von Ossietzky University Oldenburg, Ammerländer Heerstraße 114--118, 26129 Oldenburg (Oldb), Germany. ZOE, Institute for Future-Fit Economies, Bonn, Germany. oliver.richters@uni-oldenburg.de.}

\usepackage{amsmath}
\usepackage{amssymb,amsthm,amsfonts,textcomp}
\usepackage{color, transparent}
\usepackage{dpfloat}
\usepackage{array}
\usepackage{txfonts}
\usepackage{booktabs, multirow}
\usepackage{paralist}
\usepackage{hhline}
\usepackage{graphicx}
\usepackage{hanging}
\usepackage{bm} \usepackage{bbold}
\usepackage[figuresleft]{rotating}
\usepackage{acronym}
\usepackage{microtype}
\usepackage[hang]{footmisc}
\usepackage{changepage} 
\usepackage{appendix}
\usepackage{csquotes}
\usepackage{pdfpages}
\usepackage{xcolor}
\usepackage{setspace}

\usepackage{pdflscape}

\usepackage{rotating} 

\usepackage{pifont} 

\usepackage[backend=biber, natbib=true, style=authoryear-comp, sorting=nyt, url=true, isbn=false, doi=true, eprint=true, maxcitenames=2, citetracker=true, uniquename=false, labeldate=year]{biblatex}

\DeclareSourcemap{
 \maps{
  \map{
   \step[fieldsource=language, fieldset=langid, origfieldval, final]
   \step[fieldset=language, null]
  }
 }
}

\AtEveryBibitem{%
   \clearfield{day}%
   \clearfield{month}
   \clearfield{endday}%
   \clearfield{endmonth}%
}

\usepackage[inner=4.cm,outer=4.cm,top=1.2cm,bottom=0.8cm,includeheadfoot , heightrounded]{geometry}

\usepackage{chngcntr}

\makeatletter
\newcommand\arraybslash{\let\\\@arraycr}
\makeatother

\makeatletter
\g@addto@macro\UrlBreaks{\do\*\do\~\do\'\do\"\do\a\do\b\do\c\do\d\do\e\do\f\do\g\do\h\do\i\do\j\do\k\do%
\l\do\m\do\n\do\o\do\p\do\q\do\r\do\s\do\t\do\u\do\v\do\w\do\x\do\y\do\z\do\&\do\1\do\2\do\3\do\4\do\5\do\6\do\7\do\8\do\9\do\0\do\.}
\makeatother

\usepackage{scrlayer-scrpage}

\clearscrheadfoot
\ihead{\pagemark}
\ohead{\scriptsize Richters: Modeling the out-of-equilibrium dynamics of bounded rationality and economic constraints}
\let\OLDthebibliography\thebibliography
\renewcommand\thebibliography[1]{
 \OLDthebibliography{#1}
 \setlength{\parskip}{0pt}
 \setlength{\itemsep}{0pt plus 0.3ex}
}

\KOMAoption{listof}{leveldown}

\usepackage{ragged2e}
\newcolumntype{L}[1]{>{\raggedright\arraybackslash}p{#1}} 
\newcolumntype{C}[1]{>{\centering\arraybackslash}p{#1}} 
\newcolumntype{R}[1]{>{\raggedleft\arraybackslash}p{#1}} 

\newcommand{\tim}{\textcolor{gray}{{\scriptscriptstyle (t)}}}
\newcommand{\timo}{\textcolor{gray}{{\scriptscriptstyle \,}}}
\newcommand{\eq}{\textcolor{gray}{{\scriptscriptstyle \,}}}

\usepackage[hidelinks, breaklinks=true, pdfauthor={\dpautorenclean}, pdftitle={\dptitleclean}, unicode]{hyperref}

\begin{document}

\selectlanguage{english}

\thispagestyle{scrplain}

\begin{center}
 { \Large { \bfseries \sffamily \dptitle \\ \bigskip  \par \par } \vspace{1em} {\large \dpautoren \par \vspace{1em} \normalsize \dpaffiliation \par \vspace{1em} June 2021 \par } }
\end{center}
\vspace{1.3em}
\begin{addmargin}{0.03\textwidth}

\textbf{Abstract:}
The analogies between economics and classical mechanics can be extended from constrained optimization to constrained dynamics by formalizing economic (constraint) forces and economic power in analogy to physical (constraint) forces in Lagrangian mechanics.
In the differential-algebraic equation framework of General Constrained Dynamics (GCD), households, firms, banks, and the government employ forces to change economic variables according to their desire and their power to assert their interest.
These ex-ante forces are completed by constraint forces from unanticipated system constraints to yield the ex-post dynamics.
The flexible out-of-equilibrium model can combine Keynesian concepts such as the balance sheet approach and slow adaptation of prices and quantities with bounded rationality (gradient climbing) and interacting agents discussed in behavioral economics and agent-based models.
The framework integrates some elements of different schools of thought and overcomes some restrictions inherent to optimization approaches, such as the assumption of markets operating in or close to equilibrium.
Depending on the parameter choice for power relations and adaptation speeds, the model nevertheless can converge to a neoclassical equilibrium, and reacts to an austerity shock in a neoclassical or post-Keynesian way.

\bigskip

\noindent \textbf{Keywords:} Simultaneous Equation Models; Stability of Equilibrium; Balance Sheet Approach; Constrained Dynamics; Out-of-equilibrium Dynamics, Lagrangian mechanics.

\bigskip

\noindent  \textbf{JEL:} 
  A12
; B13
; C30
; C62
; E10
; E70
.

\bigskip

\noindent An earlier version of this paper was published as: Oldenburg Discussion Papers in Economics 429, March 2020, \url{https://hdl.handle.net/10419/214890}.

\bigskip

\noindent\begin{minipage}[t]{0.84\linewidth}
\textbf{Licence:} Creative-Commons \href{http://creativecommons.org/licenses/by-nc-nd/4.0/}{CC-BY-NC-ND 4.0}. \end{minipage} 
\begin{minipage}[t]{0.15\linewidth}\vspace{-\ht\strutbox}
\includegraphics[width=\columnwidth]{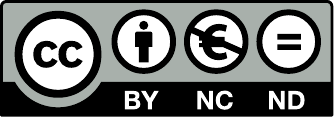}\end{minipage}

\end{addmargin}

\bigskip
\bigskip

\section*{Acknowledgments}

I thank Erhard Glötzl, Florentin Glötzl, Hans-Michael Trautwein, three reviewers and the editors for their helpful comments and acknowledge financial support from Ev.\ Studienwerk Villigst.
All errors and omissions are my own.
Conflicts of interest: none.

\pagestyle{scrheadings}

\section{Introduction}
Dynamic economic models have to describe the time evolution of stocks, flows and other variables subject to economic constraints.
The schools of economic thought differ in their modeling assumptions about rationality, heterogeneity and adaptation speeds within the economy.
The General Constrained Dynamics (GCD) framework tries to bridge some methodological gaps between these approaches and incorporate aspects from general equilibrium theory, Keynesian disequilibrium and behavioral economics (Section \ref{sec_litreview}).

Introduced by \citet{glotzl_constrained_2019}, GCD models extend the historical analogies between general equilibrium models and Newtonian physics:
Similar to the forces of interacting \emph{bodies} under constraint from Lagrangian mechanics, the modeling approach depicts the economy from the perspective of economic forces and economic power.
Economic force corresponds to the desire of agents to change certain variables, while economic power captures their ability to assert their interest to change them.
Equilibrium and optimization are replaced by out-of-equilibrium dynamics and agents gradually improving their utility in line with bounded rationality discussed in behavioral economics.
The introduction of constraint forces, i.\,e.\ forces arising from system constraints, allows for a consistent assessment of ex-ante and ex-post dynamics of the dynamical system.

The conceptual model presented in Section \ref{sec_model} is the first GCD model that incorporates two household sectors, two production sectors with input--output relations, banks and the government.
It allows to integrate aspects from general equilibrium, Keynesian disequilibrium and behavioral economics resp. agent-based models simultaneously.
It is based on a Keynesian balance sheet approach in which quantities adjust gradually and prices react slowly on supply-demand mismatches.
The stability analysis in Section \ref{model-analysis} reveals the conditions and power relations under which convergence to the usual neoclassical equilibrium is achieved.
Fast adaptation of quantities and prices does not lead to fast convergence, but can amplify deviations from the equilibrium.
Depending on parameters, the model reacts to fiscal consolidation in a neoclassical or post-Keynesian way.
Conspicuous consumption can be integrated as mutual influence on consumption decisions.
Section \ref{sec_conclusion} concludes.

\section{Modeling dynamics subject to constraints in different schools of thought and in GCD models}
\label{sec_litreview}
\label{sec_framework}

General Constrained Dynamics \citep{glotzl_why_2015, glotzl_constrained_2019} is an economic framework, describing the interaction of bounded rational agents that exert economic forces to improve their situation (gradient climbing) subject to the economic constraints.
Adapting the concept of constrained dynamics from Lagrangian mechanics \citep{lagrange_mechanique_1788, flannery_dalembertlagrange_2011}, it carries on an ``unfinished business'' \citep[pp.~26--30]{leijonhufvud_episodes_2006} of the early neoclassicals such as Irving Fisher \citeyearpar{fisher_mathematical_1892} or Vilfredo Pareto \citeyearpar{pareto_cours_1897}:
Inspired by the description of stationary states in classical mechanics, they derived an economic theory of static equilibrium \citep{pikler_utility_1955, mirowski_more_1989, grattan-guinness_how_2010, glotzl_constrained_2019}.
Despite some efforts, they were unable to describe analogously the adaptive \emph{processes} that were thought to converge to the states analyzed in \emph{static} theory \citep{donzelli_paretos_1997, mclure_pareto_2001, leijonhufvud_episodes_2006}.
The GCD approach aims at closing this gap.

In the following, we review how economic models treat constraints, behavioral assumptions, and their compatibility -- and compare it to the GCD approach.
In general, each dynamic economic model is described by $J$ agents and $I$ time-dependent variables $x_i(t)$ that can correspond to any stocks or flows of commodities, resources, financial liabilities, or any other variables or parameters such as prices or interest rates.

\subsection{Constraints in economic models}

The \emph{structure} of each model consists of $K$ economic constraints that remove many degrees of freedom.
Constraints can be identities, relations ``that hold by definition'' \citep[p.~4]{allen_macro-economic_1982} such as the national income account identity or accounting constraints in balance sheets.
In material flow analysis \citep{brunner_practical_2004}, constraints include laws of nature such as conservation of mass and energy as \emph{first laws} of chemistry and thermodynamics.
Input--output relations or production functions imply certain technological limitations, while budget constraints are derived from the behavioral assumption that nobody is giving away money without an equivalent remuneration.
The respect for identities is ``the beginning of wisdom'' in economics, but they must not be ``misused to imply causation'' \citep[p.~11]{tobin_policies_1995}.
To derive causal arguments, a \emph{closure} has to be chosen that combines individual agency and the constraints:
If the $I$ variables were influenced by $I$ behavioral equations, the system of equations would be overdetermined because of the additional $K$ constraints.
The schools of economic thought differ in their ways of making this system of equations solvable \citep{taylor_income_1991}, depending on the behavioral assumptions.
In the GCD approach, constraints are implemented as algebraic equations that typically depend on variables and their time derivatives:
\begin{align}
0 &= Z_k(x, \dot x), \quad k \in \{1,\ldots, K\}.
\end{align}

\subsection{Behavior in economic models}

In most general equilibrium models, each agent is assumed to fully control and voluntarily adapt all the stocks and flows directly affecting him (such as individual working hours or savings).
The optimization of an (intertemporal) utility function subject to the constraints results in various individual first-order conditions.
To solve a macroeconomic model, this society of utility maximizers has to be aggregated into a single function -- because every optimization approach requires \emph{one single} function to be optimized.
The assumption of a ``representative agent'' sidesteps this problem by assuming all agents to be identical \citep[p.~350]{blundell_heterogeneity_2005}.

Equilibrium models that include heterogeneity among households and firms \citep{kaplan_monetary_2016, christiano_dsge_2018} have to limit heterogeneity and social influences to allow for aggregation.
Unfortunately, assumptions made about individual rationality are ``not enough to talk about social regularities'',
but it is necessary that ``macro-level assumptions \ldots\ restrict the \emph{distribution} of preferences or endowments'' to guarantee a unique equilibrium \citep[p.~359--63]{rizvi_microfoundations_1994}.
Aggregation is possible if and only if demand is independent of the distribution of income among the agents \citep{gorman_class_1961, stoker_empirical_1993, kirman_market_1986, kirman_whom_1992}, which \citet[p.~363]{rizvi_microfoundations_1994} calls an ``extremely special situation''.

Behavioral economics and agent-based models (ABM) assume that individuals cannot solve infinitely dimensional optimization problems, but use bounded rationality instead.
They emphasize that interactions between heterogeneous agents matter beyond market prices, and social interaction, social norms, power relations or institutions influence economic choices.
Compared to selfish utility maximizers, this corresponds to a broader version of methodological individualism,
because the aggregate dynamics cannot be deduced from individual behavior.
In many ABM and post-Keynesian models, agents follow a sequence of simple rule-of-thumb behavior instead of an optimization procedure \citep{gallegati_agent_2009, schoder_keynesian_2020, godley_monetary_2012}.
\citet[p.~248]{lindenberg_social_2001} describes bounded rationality as the ``general desire to improve one’s condition'', and \citet[p.~244]{munier_bounded_1999} discuss ``procedural optimizing'' as possible modeling strategy.

GCD models formalize these ideas as agents that try to \emph{slowly increase} their utility with a gradient climbing approach.
Each agent seeks to improve the existing configuration in the direction of his desires:
The dynamics of the model are the result of \emph{economic forces} and \emph{economic power}:
An economic force $f_{ji}$ corresponds to the desire of agent $j$ to change a certain variable $x_i$. Economic power $\mu_{ji}$ captures the ability of an agent $j$ to assert its interest to change variable $x_i$.\footnote{The economic power factors $\mu_{ji}$ as \emph{ability to change} a variable correspond to the inverse of the mass in the Newtonian equations, in which mass is the \emph{resistance} to a change of velocity (\citealp[p.~382]{estola_newtonian_2017}; \citealp{glotzl_constrained_2019}).}
The total impact on the variable $x_i$ is the product of economic force and power $\mu_{ji} f_{ji}$, i.\,e.\ the product of \emph{desire} and \emph{ability}:
\begin{align}
\dot x_i(t) &= \sum_{j=1}^J \mu_{ji} f_{ji}(x). \label{eq_generalmodel}
\end{align}
All agents are unable to calculate infinite dimensional intertemporal optimization problems based on rational expectations about the reactions of the other market participants.
Instead, they base their decisions on how much to work, invest, consume or save on the observation of current marginal utilities, profits, productivities and prices.
They do not jump to the point of highest utility as rational utility \emph{maximizers}, but instead try to ``climb up the utility hill'' gradually by pushing the economy in the direction of highest marginal utility.
In a continuous time framework, this can be modeled by defining the forces exerted by the agents as gradients of their utility functions.
With this gradient seeking approach, agents still satisfy the definition of rationality by \citet[p.~6]{mankiw_principles_2008}:
``A rational decision maker takes an action if and only if the marginal benefit of the action exceeds the marginal cost.''
One might say that the agents in the economy are as rational as shortsighted first year economics students.

\subsection{Making behavior and constraints consistent}

Economic models face the challenge of making the behavioral decisions consistent with the constraints.
In general equilibrium models, satisfying the $K$ system constraints of market exchange can only be guaranteed by letting $K$ prices adapt that make all the individual plans compatible with each other (market clearing as neoclassical closure).
Interacting via price signals, constraints imposed by other agents or system properties can be fully anticipated by the agents \citep{arrow_general_1971}.
Therefore, equilibrium models do not allow to study ``the emergence and propagation of macroeconomic inconsistencies'' \citep[p.~5]{guzman_towards_2020}.

Keynesian disequilibrium models assume that price adaptations cannot clear markets sufficiently fast.
This implies that the \emph{ex-ante} (planned) behavior does not necessarily respect the economic constraints, and the \emph{ex-post} (actual) dynamics are influenced by both system constraints and the agency of others:
Terms such as ``forced saving'' or ``involuntary unemployment'' \citep{barro_general_1971} indicate that agents cannot have complete control over the variables affecting them.
For example, in some Keynesian disequilibrium models, quantities of voluntary exchange are rationed by the \emph{short-side}: Depending on market conditions, demand is limited by insufficient supply or otherwise \citep{benassy_neo-keynesian_1975, malinvaud_theory_1977}.
Other models consider the labor market to be demand-led and employees have \emph{no} influence on working times.
For example, \citet{schoder_keynesian_2020} treats the nominal wage as a policy variable and allows for unemployment in a Keynesian Dynamic Stochastic Disequilibrium model.
For each price set exogenously, one quantity \emph{has} to be \emph{not} determined by market clearing in order to avoid an overdetermined system of equations.
In accordance with this, the $K$ constraints that guarantee consistency in post-Keynesian Stock-Flow Consistent models are satisfied by simply dropping $K$ behavioral equations \citep{godley_monetary_2012, caverzasi_post-keynesian_2015}.
This one-sided \emph{drop closure} is justified if and only if exactly $K$ stocks or flows are unaffected by agency, but only determined by the constraints \citep[for a critique, see][]{richters_modeling_2020}.

Agent-Based Models focus on the interaction of bounded rational agents and the emergence of either a (statistical) equilibrium, but also discontinuities, tipping points, lock-ins or path dependencies \citep{kirman_economic_2010}.
ABM lack a common core, and different coordinating mechanisms such as price adaptations, auctions, matching algorithms or quantity rationing are implemented to account for the economic constraints \citep{tesfatsion_agent-based_2006, gintis_dynamics_2007, gallegati_agent_2009, lengnick_agent-based_2013, ballot_agent-based_2014}.
The evolution of stocks and flows in some ABM was logically incoherent, for example because defaulted firms were simply recapitalized ``ex-nihilo''.
Stock-flow consistent agent-based models avoid these inconsistencies \citep{caiani_agent_2016, caverzasi_toward_2018}.

To guarantee consistency, GCD models use a ``Lagrangian closure'' based on analogies to constraint forces in physics \citep{glotzl_constrained_2019}:\footnote{The ``Newtonian Microeconomics'' approach by \citet{estola_testing_2012} and \citet{estola_newtonian_2017} is similar in the formalization of economic forces, but they accept that supply and demand differ not only ex-ante, but also ex-post \citep[pp.~222, 386]{estola_newtonian_2017}.
This violation of economic identities occurs because they lack a formalization of economic constraint forces.}
If all the variables $x_i$ in a constraint $Z_k$ are affected by agency, additional constraint forces $z_{ki}$ are added to the time evolution of $x_i$, which together with the forces $f_{ji}$ applied by all agents with power factors $\mu_{ji}$ creates the ex-post dynamics:
\begin{align}
\dot x_i(t) &= \sum_{j=1}^J \mu_{ji} f_{ji}(x) + \sum_{k=1}^K z_{ki}(x, \dot x),  \label{eq_generalmodel-constraint} \\
0 &= Z_k(x, \dot x).
\end{align}
The constraint forces lead to unintended deviations of the actual time evolution from the planned one, such that the emerging behavior is consistent with all constraints.
In economics, the magnitude of the constraint forces $z_{ki}$ cannot be derived from laws of nature, but reflect assumptions about adaptation processes within the economic system.
In physics \citep{flannery_dalembertlagrange_2011, glotzl_constrained_2019}, the time-dependent constraint forces $z_{ki}$ can be calculated as
\begin{align}
    z_{ki}(x,\dot x) &= \lambda_k \frac{\partial Z_k}{\partial x_i}, \label{eq_economics_3}
\end{align}
or, if $\partial Z_k / \partial x_i \equiv 0$, as
\begin{align}
    z_{ki}(x,\dot x) &= \lambda_k \frac{\partial Z_k}{\partial \dot x_i}. \label{eq_economics_4}
\end{align}
The additional variable $\lambda_k$ (`Lagrangian multiplier') is introduced to make the model solvable.
This rule from mechanics is a plausible choice also in economics \citep{glotzl_constrained_2019}, and the static version of these constraint forces is known from optimization exercises such as maximizing $U(x_1, x_2)$ subject to a budget constraint $0 = M - p_1 x_1 - p_2 x_2$.
The first order condition $0 = \frac{\partial U}{\partial x_1} - \lambda p_1$ means that the utility force and the constraint force cancel out, the latter given by the derivative of the constraint with a Lagrangian multiplier $\lambda$ similar to Eqs.~(\ref{eq_economics_3}--\ref{eq_economics_4}).
The gradient climbing approach and the constraint forces in analogy to classical mechanics are the reason why the GCD approach has to be formulated in continuous time.
The system of differential-algebraic equations (Eqs.~\ref{eq_generalmodel-constraint}--\ref{eq_economics_4}) can be solved numerically for $x(t)$ and $\dot x(t)$.

\section{The model}
\label{sec_model}

Introducing the GCD framework, \citet{glotzl_constrained_2019} presented a microeconomic Edgeworth box exchange model with two agents and two commodities and slow price adaptation that converges to the neoclassical contract curve for most parameters. 
\citet{richters_modeling_2020} described a simple post-Keynesian stock-flow consistent disequilibrium model of the macroeconomic monetary circuit in this framework.
This paper extends these ideas to a complex macroeconomic model that includes production, real capital, inventories, and two production sectors with intermediate goods and two household sectors.
The conceptual model is designed to show how the simultaneous processes of trade by bounded rational agents and slow price adaptation can converge to a conventional general equilibrium solution.
It integrates some aspects important for agent-based models and Keynesian disequilibrium models, but because of the limited number of agents, it lacks complex dynamics of interaction, evolution and emergence.

\subsection{Model structure: the constraints}

\begin{figure*}[tp]
\hfil\includegraphics[width=1\textwidth]{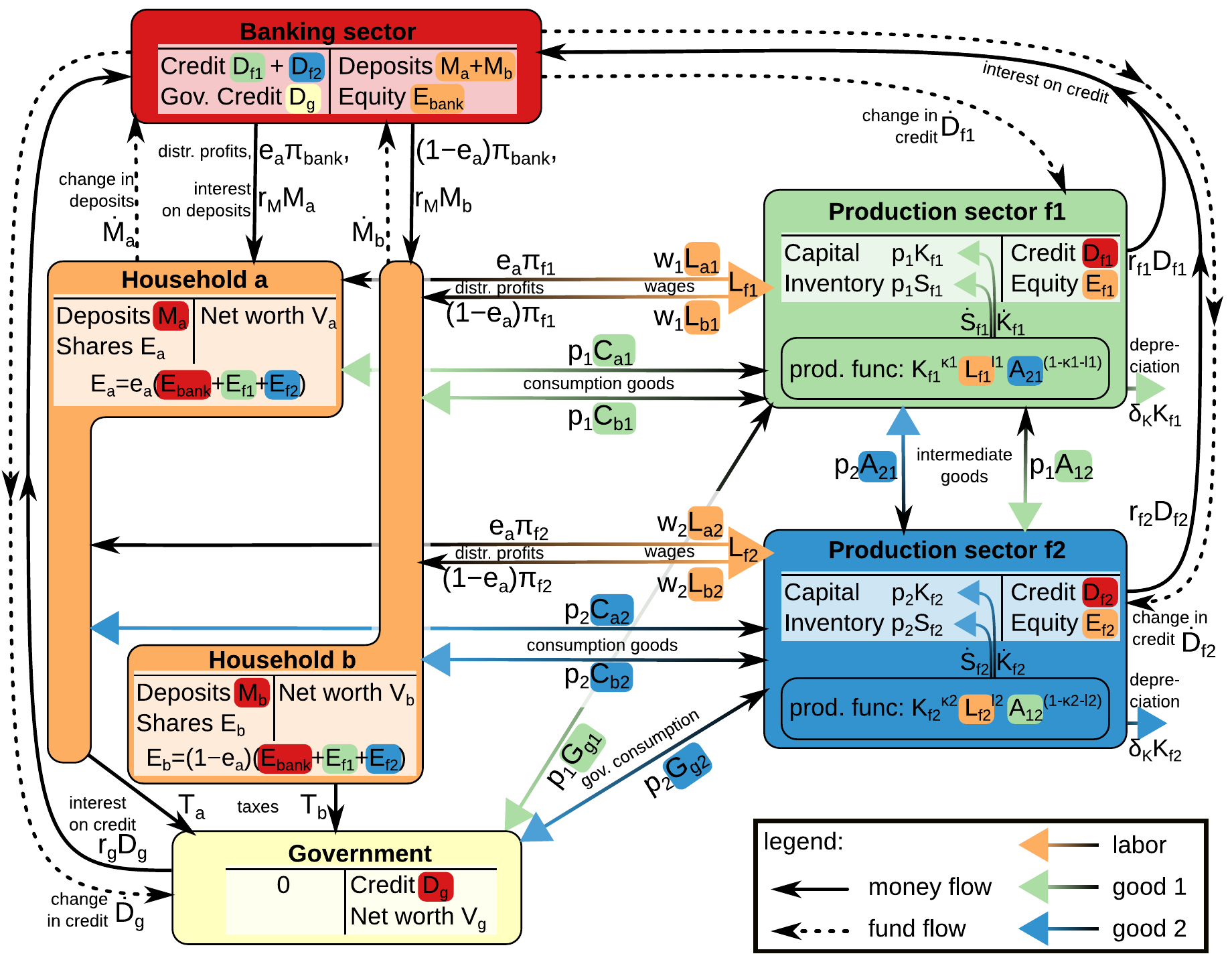}\hfil
\caption{\label{fig:model-structure}Model structure: The diagram depicts the balance sheets of the different sectors and the flows of money (black arrows), funds (dotted arrows) and goods (colored arrows) within the economy. The interconnectedness of the balance sheets is depicted by background colors: For example, the liability of production sector $f1$ towards the bank has a red background, while the corresponding claim in the bank's balance sheet has a green background. \newline
The six balance sheets provide the constraints in Eqs.~(\ref{balance-sheet-f1}--\ref{balance-sheet-g}).
Consistency of money flows provides the budget constraints (Eqs.~\ref{model-budget-a}--\ref{model-budget-bank}).
Eqs.~(\ref{model-structure-L1}--\ref{model-structure-L2}) reflect consistency of labor flows, while consistency of the flows of good 1 and good 2 provides Eqs.~(\ref{model-firm-prod1}--\ref{model-firm-prod2}).
}
\end{figure*}

\begin{table}[tp]
{\footnotesize
\begin{tabular}{lccccc}
\toprule
& Household sector $h$ & Prod. sector $fi$ & Banks & Gov. & Total \\\midrule
Fixed capital & & $+ p_i K_{fi}$ & & & $+\sum\nolimits_i p_i K_{fi}$ \\
Inventories & & $+ p_i S_{fi}$ & & & $+\sum\nolimits_i p_i S_{fi}$ \\
Deposits & $+M_h$ & & $-\sum\nolimits_h M_h$ & & 0 \\
Credit & & $-D_{fi}$ & $+D_g + \sum\nolimits_i D_{fi}$ & $-D_g$ & 0 \\
Equity/Shares & $+e_h (E_{bank} + \sum\nolimits_i E_{fi})$ & $-E_{fi}$ & $- E_{bank}$ & & 0 \\ \midrule
Balance/Net worth & $-V_h$ & 0 & 0 & $-V_g$ & $-\sum\nolimits_i p_i (K_{fi} + S_{fi})$ \\ \midrule
Sum & 0 & 0 & 0 & 0 & 0 \\\bottomrule
\end{tabular}
}
\caption{\label{tab:balance-sheet} Nominal balance sheets. Household sectors are indexed with $h$ ($\sum e_h = 1$), production sectors with $i$. A $+$ before a magnitude denotes an asset, whereas $-$ denotes a liability. For details, see \citet{nikiforos_stock-flow_2017}.
}
\end{table}

The model studies the interaction of two household sectors, two production sectors, a bank, and the government.
They trade two consumer goods, labor and capital, financed by bank credit or equity.
Firms are privately owned and equity is valued at book value.
All agents show bounded rationality and try to increase their utility with a gradient climbing approach.
Prices react slowly on demand--supply mismatches.

The balance sheet and the transaction matrix can be found in Tables \ref{tab:balance-sheet} and \ref{tab:transaction-matrix} (on page \pageref{tab:transaction-matrix}).
To avoid the flaw of inconsistent accounting present in some monetary circuit models \citep[see the critique by][]{zezza_godley_2012, richters_consistency_2017}, they allow to track the consistency of stocks and flows.

\begin{sidewaystable}[p]
{\footnotesize \begin{tabular}{lcccccccc}
\toprule
 & \multirow{2}{*}{\textbf{Household sector $h$}} & \multicolumn{2}{c}{\textbf{Production sector $fi$}} & \multicolumn{2}{c}{\textbf{Banks}} & \multirow{2}{*}{\textbf{Government}} & \multirow{2}{*}{\textbf{$\sum$}} \\
 &  & \textbf{Current Account} & \textbf{Capital Acc.} & \textbf{Current Acc.} & \textbf{Capital Acc.} &  &  &  \\ \midrule
\textbf{Transactions} &  &  &  &  &  &  &  &  \\ 
Consumption & $-\sum\nolimits_i p_i C_{hi}$ & $+\sum\nolimits_h p_i C_{hi}$ &  &  &  &  & 0 &  \\ 
Govt. Spending &  & $+p_i G_{gi}$ &  &  &  & $-\sum\nolimits_i p_i G_{gi}$ & 0 &  \\ 
Taxes & $-T_h$ &  &  &  &  & $+\sum\nolimits_h T_h$ & 0 &  \\ 
Wages & $+\sum\nolimits_i w_i L_{hi}$ & $-\sum\nolimits_h w_i L_{hi}$ &  &  &  &  & 0 &  \\ 
Investment into Capital &  & $+ p_i (\dot K_{fi} + \delta_K K_{fi})$ & $- p_i (\dot K_{fi} + \delta_K K_{fi})$ &  &  &  & 0 &  \\ 
Investment into Inventory &  & $+ p_i \dot S_{fi}$ & $- p_i \dot S_{fi}$ &  &  &  & 0 &  \\ 
Depreciation allowances &  & $- p_i \delta_K K_{fi}$ & $+ p_i \delta_K K_{fi}$ &  &  &  & 0 &  \\ 
Interest on credit &  & $-r_{fi} D_{fi}$ &  & $r_g D_g + \sum\nolimits_i r_{fi} D_{fi}$ &  & $-r_g D_g$ & 0 &  \\ 
Interest on deposits & $+r_M M_h$ &  &  & $-\sum_h r_M M_h$ &  &  & 0 &  \\ 
Distributed profits & $+e_h (\Pi_{bank} + \sum\nolimits_i \Pi_{fi})$ & $-\Pi_{fi}$ &  & $-\Pi_{bank}$ &  &  & 0 &  \\ 
Intermediate Purch. &   & $\sum\nolimits_{j\neq i} (p_i  A_{ij} - p_j A_{ji})$ &  &  &  &  & 0 &  \\ \midrule
\textbf{Flow of Funds} &  &  &  &  &  &  &  &  \\ 
Change in deposits & $-\dot M_h$ &  &  &  & $+ \sum\nolimits_h \dot M_h$ &  & 0 &  \\ 
Change in credit &  &  & $+\dot D_{fi}$ &  & $-\dot D_g -\sum\nolimits_i \dot D_{fi}$ & $+\dot D_g$ & 0 &  \\ 
Change in capital &  &  & $+ p_i \dot K_{fi}$ &  &  &  & $+ p_i \dot K_{fi}$ &  \\ 
Change in inventory &  &  & $+ p_i \dot S_{fi}$ &  &  &  & $+ p_i \dot S_{fi}$ &  \\ \midrule
Revaluation of Capital &  &  & $+\dot p_i K_{fi}$ &  &  &  & $+\dot p_i K_{fi}$ &  \\ 
Revaluation of Inventories &  &  & $+\dot p_i S_{fi}$ &  &  &  & $+\dot p_i S_{fi}$ &  \\ 
Revaluation of Shares & $-e_h (\dot E_{bank} + \sum_i \dot E_{fi})$ &  & $+\dot E_{fi}$ &  & $+\dot E_{bank}$ &  & 0 &  \\ 
Change in net worth & $\dot V_h$ & 0 & 0 & 0 & 0 & $\dot V_g$ & $\partial_t (\sum_i p_i (K_{fi} + S_{fi}))$ &  \\ \bottomrule
\end{tabular} }
\caption{\label{tab:transaction-matrix}The matrix tabulates all transactions and flows of funds \citep[see][]{nikiforos_stock-flow_2017}.}
\end{sidewaystable}

As depicted in Figure~\ref{fig:model-structure}, the model consists of $42$ economic variables:
\begin{itemize}
\setlength{\itemsep}{0pt}
\item 11 financial balance sheet entries: $M_a$, $M_b$, $V_a$, $V_b$, $E_{bank}$, $E_{f1}$, $E_{f2}$, $D_{f1}$, $D_{f2}$, $D_g$, $V_g$,
\item 4 stocks of real capital and inventories: $K_{f1}$, $K_{f2}$, $S_{f1}$, $S_{f2}$,
\item 8 prices: $r_{f1}$, $r_{f2}$, $r_g$, $r_M$, $p_1$, $p_2$, $w_1$, $w_2$,
\item 6 flows of labor: $L_{a1}$, $L_{a2}$, $L_{b1}$, $L_{b2}$, $L_{f1}$, $L_{f2}$,
\item 8 flows of goods: $C_{a1}$, $C_{a2}$, $C_{b1}$, $C_{b2}$, $G_{g1}$, $G_{g2}$, $A_{12}$, $A_{21}$,
\item 5 flows of money: $\pi_{f1}$, $\pi_{f2}$, $\pi_{bank}$, $T_a$, $T_b$.
\end{itemize}
They are related by $16$ constraints (Eqs.~\ref{balance-sheet-f1}--\ref{model-firm-prod2}).

The consistency of double-entry bookkeeping in each of the six sectors as shown in Table \ref{tab:balance-sheet} provides six mathematical constraints:
\renewcommand{\tim}{{}}
\begin{align}
 0 &= p_1\tim (K_{f1}\tim + S_{f1}\tim) - D_{f1}\tim        - E_{f1}\tim,  \label{balance-sheet-f1}  \\
 0 &= p_2\tim (K_{f2}\tim + S_{f2}\tim) - D_{f2}\tim        - E_{f2}\tim,  \label{balance-sheet-f2} \\
 0 &= D_{f1}\tim + D_{f2}\tim + D_g\tim - M_a\tim - M_b\tim - E_{bank}\tim, \label{balance-sheet-bank} \\
 0 &= M_a\tim + e_a(E_{f1}\tim + E_{f2}\tim + E_{bank}\tim) - V_a\tim, \\
 0 &= M_b\tim + (1-e_a) (E_{f1}\tim + E_{f2}\tim + E_{bank}\tim) - V_b\tim, \\
 0 &= 0 - D_g\tim                                           - V_g\tim. \label{balance-sheet-g}
\end{align}
Final goods, capital and inventories are assumed to have the same price.
The balance sheets are interconnected, because every financial claim has a corresponding liability, depicted by the colored background of the entries in Figure~\ref{fig:model-structure}.
Household sector $a$ holds a fraction $e_a$ of the shares of the firm and banking sectors, while $1-e_a$ is left for household sector $b$.
They cannot trade their stakes in the firms.
Eqs.~(\ref{balance-sheet-f1}--\ref{balance-sheet-g}) are used as definitions for $E_{f1}$, $E_{f2}$, $E_{bank}$, $V_a$, $V_b$ and $V_g$.
Therefore, no Lagrangian multipliers are needed to guarantee consistency.
Summing all these equations yields $V_a + V_b + V_g = p_1\tim (K_{f1}\tim + S_{f1}\tim) + p_2\tim (K_{f2}\tim + S_{f2}\tim)$, thus the actual wealth consists of real stocks of capital and inventories, because the credit relations cancel out.
In the following, the equations for household sector $b$ and production sector $f2$ are provided, but explanations refer to household sector $a$ and sector $f1$ only.

Six budget constraints track the flow of money for each agent.
Household sector $a$ consumes an amount $C_{a1}$ at price $p_1$ from sector $f1$ and $C_{a2}$ at price $p_2$ from sector $f2$.
It works an amount $L_{a1}$ for wage $w_1$ in sector $f1$ and $L_{a2}$ for wage $w_2$ in sector $f2$, but has to pay taxes, for simplicity only on labor income, with an exogenous tax rate $\theta$.
Additional to wages, it receives a share $e_a$ of the total distributed profits of production sectors and banks, while the deposits $M_a$ earn him a yearly interest of $r_M M_a$.
The budget constraints are:
\begin{align}
\begin{split}
     Z_a\tim = 0 &= \dot M_a\tim + p_1\tim C_{a1}\tim + p_2\tim C_{a2}\tim -\left(1-\theta\right) \left(w_1\tim L_{a1}\tim +w_2\tim L_{a2}\tim \right) \\
& \qquad \ldots - e_a(\pi_{f1}\tim + \pi_{f2}\tim +\pi_{bank}\tim) - r_M\tim M_a\tim, \label{model-budget-a} \end{split}\\
\begin{split}
     Z_b\tim = 0 &= \dot M_b\tim + p_1\tim C_{b1}\tim + p_2\tim C_{b2}\tim -\left(1-\theta\right) \left(w_1\tim L_{b1}\tim +w_2\tim L_{b2}\tim \right) \\
& \qquad \ldots - (1-e_a)(\pi_{f1}\tim + \pi_{f2}\tim +\pi_{bank}\tim) - r_M\tim M_b\tim. \label{model-budget-b} \end{split}
\end{align}
The government $g$ pays interest $r_g$ on government credit $D_g$ and buys goods $G_{g1}$ and $G_{g2}$ at price $p_1$ and $p_2$ from the two production sectors.
It levies wage taxes with a constant tax rate $\theta$, which results in the following budget constraint:
\begin{align}
     Z_g\tim = 0 &= p_1\tim G_{g1}\tim + p_2\tim G_{g2}\tim -\theta \left( w_1\tim L_{a1}\tim + w_2\tim L_{a2}\tim + w_1\tim L_{b1}\tim + w_2\tim L_{b2}\tim \right) + r_g\tim D_g\tim - \dot D_g\tim. \label{model-budget-g}
\end{align}
Production sector $f1$ (and equivalently $f2$) has to pay a wage $w_1$ per unit of work, a price $p_2$ for intermediate goods $A_{21}$ used in production, and interest $r_{f1}$ on credit $D_{f1}$.
Money inflows arise from selling goods at price $p_1$ to households, the government, and sector $f2$.
The difference between money inflows and outflows is distributed as profits $\pi_{f1}$ or changes the stock of credit $\dot D_{f1}$, implying the following budget constraints:
\begin{align}
  Z_{f1}\tim = 0 &= w_1\tim L_{f1}\tim + p_2\tim A_{21}\tim + r_{f1}\tim D_{f1}\tim - p_1\tim (C_{a1}\tim  + C_{b1}\tim  + G_{g1}\tim + A_{12}\tim ) + \pi_{f1}\tim - \dot D_{f1}, \label{model-budget-f1} \\
  Z_{f2}\tim = 0 &= w_2\tim L_{f2}\tim + p_1\tim A_{12}\tim + r_{f2}\tim D_{f2}\tim - p_2\tim (C_{a2}\tim  + C_{b2}\tim  + G_{g2}\tim + A_{21}\tim ) + \pi_{f2}\tim - \dot D_{f2}. \label{model-budget-f2}
\end{align}
The banking sector receives interest payments on credits and pays interest $r_M M_a$ and $r_M M_b$ to households.
The difference between money inflows and outflows is distributed as profits $\pi_{bank}$ or changes the stock of equity $\dot E_{bank}$, implying the following budget constraint:
\begin{align}
Z_{bank}\tim = 0 &= r_M\tim \left(M_a\tim +M_b\tim \right) -  r_{f1}\tim D_{f1}\tim - r_{f2}\tim D_{f2}\tim - r_g\tim D_g\tim + \pi_{bank}\tim + \dot E_{bank}\tim. \label{model-budget-bank}
\end{align}

Note that the constraints $Z_a$, $Z_b$, $Z_g$, $Z_{f1}$, $Z_{f2}$ and $Z_{bank}$ are linearly dependent with the time derivative of Eq.~\eqref{balance-sheet-bank} -- as in every stock-flow consistent model, one budget constraint is redundant \citep[p.~395]{godley_money_1999}. Consequently, Eq.~\eqref{balance-sheet-bank} can be dropped and will just serve as an initial condition for $t = 0$, resulting in $15$ linearly independent constraints.

Labor input $L_{f1}$ of sector $f1$ has to be identical to the amount of work in this sector by household sectors $a$ and $b$, interrelating the variables of those agents:
\begin{align}
Z_{L1}\tim = 0 &= L_{a1}\tim + L_{b1}\tim - L_{f1}\tim, \label{model-structure-L1} \\
Z_{L2}\tim = 0 &= L_{a2}\tim + L_{b2}\tim - L_{f2}\tim. \label{model-structure-L2}
\end{align}
As households and firms influence all these variables, these constraints cannot be treated as definitions for one variable.
Consequently, constraint forces with Lagrangian multipliers $\lambda_{L1}$ and $\lambda_{L2}$ are added to the time evolution of these variables to ensure consistency (Note: the index $_i$ is identical for the Lagrangian multipliers $\lambda_i$ and the corresponding constraints $Z_i$ throughout the paper).
$\lambda_{L1}$ is negative if the desired change in variables would lead to ex-ante excess supply for labor in sector $f1$.
It will show up as constraint force in the time evolution of $L_{a1}$, $L_{b1}$ and $L_{f1}$ (Eqs.~\ref{model-household-La1}, \ref{model-household-Lb1}, \ref{model-firm-Lf1}), and the time evolution of the wage $w_1$ (Eq.~\ref{model-w1dot}).

A constraint within sector $f1$ is that total production given by a Cobb-Douglas production function depending on capital $K_{f1}$, labor $L_{f1}$ and intermediate input $A_{21}$ has to be equal to consumption by households $C_{a1} + C_{b1}$, government consumption $G_{g1}$, deliveries to sector $f2$ as intermediate goods $A_{12}$, gross investment $\delta_K K_{f1} + \dot K_{f1}$ and change in inventory $\dot S_{f1}$.
Sector $f2$ is constructed symmetrically, assuming a circular-horizontal production structure.
\begin{align}
  Z_{P1}\tim = 0 &= K_{f1}\tim^{\kappa_1} L_{f1}\tim^{l_1} A_{21}\tim^{1-\kappa_1-l_1} -\dot K_{f1}\tim -\delta_K K_{f1}\tim -C_{a1}\tim -C_{b1}\tim -G_{g1}\tim -\dot S_{f1}\tim -A_{12}\tim,   \label{model-firm-prod1} \\
  Z_{P2}\tim = 0 &= K_{f2}\tim^{\kappa_2} L_{f2}\tim^{l_2} A_{12}\tim^{1-\kappa_2-l_2} -\dot K_{f2}\tim -\delta_K K_{f2}\tim -C_{a2}\tim -C_{b2}\tim -G_{g2}\tim -\dot S_{f2}\tim -A_{21}\tim.   \label{model-firm-prod2}
\end{align}
These identities will be guaranteed by the Lagrangian multipliers $\lambda_{P1}$ and $\lambda_{P2}$.

\subsection{Agents' behavior}

Given 42 variables and 15 linearly independent constraints, only 27 behavioral equations could be chosen without the concept of Lagrangian closure.
To show the flexibility of the framework, both the Lagrangian closure and the drop closure will be used for different variables.
In the latter case, the behavior is implemented as an algebraic equation, not a differential equation.
The model considers behavioral influences on 34 variables, which results in 49 equations for 42 variables.
Therefore, 7 Lagrangian multipliers have to be added, one for each constraint in which all the variables are influenced by behavior.
The following sections explain the constraints and behavioral assumptions in detail for households, government, firms and banks.

\renewcommand{\tim}{\textcolor{gray}{{\scriptscriptstyle (t)}}}

\subsubsection{Household sectors}
\label{model-households}

The households are assumed to derive utility from consumption, leisure, and government consumption. In each variable, the utility functions $U_a$ and $U_b$ satisfy the Inada conditions:\footnote{$U$ is strictly increasing, strictly concave, continuously differentiable and $U^\prime(0) = \infty$ and $U^\prime(\infty ) = 0$ in every argument.}
\begin{align}
U_a\tim &= C_{a1}\tim^{\alpha_{C1}} C_{a2}\tim^{\alpha_{C2}} + \left(1-L_{a1}\tim -L_{a2}\tim \right)^{\alpha_L} + G_{g1}\tim^{\alpha_G}, \label{eq_utility_a} \\ 
U_b\tim &= C_{b1}\tim^{ \beta_{C1}} C_{b2}\tim^{ \beta_{C2}} + \left(1-L_{b1}\tim -L_{b2}\tim \right)^{ \beta_L} + G_{g2}\tim^{\beta_G}. \label{eq_utility_b} 
\end{align}
Note that we assume that the household sector $a$ derives utility from the government expenditure in sector $f1$, while the expenditure in sector $f2$ is relevant for household sector $b$.
The reason for this choice is to be able to model the political struggle between the two to shift government expenditure towards \emph{their} sector.

Ex-post, households' decisions must be consistent with the budget constraints (Eqs.~\ref{model-budget-a}--\ref{model-budget-b}).
The constraint forces are proportional to the Lagrangian multiplier $\lambda_a$ times the derivative of the constraint $Z_a$ with respect to the particular variable (see Section \ref{sec_framework}).

For work $L$, the derivative of the budget constraint yields $\tfrac{\partial Z_a}{\partial L_{a1}} = - (1-\theta) w_1$, $\tfrac{\partial Z_a}{\partial L_{a2}} = - (1-\theta) w_2$.
Additionally, the structural equations (Eqs.~\ref{model-structure-L1}--\ref{model-structure-L2}) for labor have to be satisfied.
To avoid that total labor in a sector is different from the sum of work performed by all the households in this sector, an additional constraint force is added.
Following the Lagrangian closure, the constraint forces are proportional to the derivative, $\tfrac{\partial Z_{L1}}{\partial L_{a1}} = \tfrac{\partial Z_{L2}}{\partial L_{a2}} = +1$ and $\tfrac{\partial Z_{L1}}{\partial L_{f1}} = -1$, which implies that all these variables are adjusted by the same amount.
If $\dot L_{a1} + \dot L_{b1} > \dot L_{f1}$ ex-ante, the constraint forces reduce $\dot L_{a1}$ and $\dot L_{b1}$ while increasing $\dot L_{f1}$ until consistency is reached.
Instead, a post-Keynesian economist may assume that firms' demand fully determines households' supply of labor, which illustrates that the choice of constraint forces reflects assumptions about power relations within the economy.
Taken together, the gradient forces from the utility function and the constraint forces yield the following time evolution:
\begin{align}
\dot L_{a1}\tim &= -\mu_{aL1} \cdot \alpha_L \left(1-L_{a1}\tim -L_{a2}\tim \right)^{\alpha_L-1} - \lambda_a\tim w_1\tim (1-\theta) + \lambda_{L1}\tim, \label{model-household-La1} \\
\dot L_{a2}\tim &= -\mu_{aL2} \cdot \alpha_L \left(1-L_{a1}\tim -L_{a2}\tim \right)^{\alpha_L-1} - \lambda_a\tim w_2\tim (1-\theta) + \lambda_{L2}\tim, \label{model-household-La2}  \\
\dot L_{b1}\tim &= -\mu_{bL1} \cdot  \beta_L \left(1-L_{b1}\tim -L_{b2}\tim \right)^{ \beta_L-1} - \lambda_b\tim w_1\tim (1-\theta) + \lambda_{L1}\tim, \label{model-household-Lb1}  \\
\dot L_{b2}\tim &= -\mu_{bL2} \cdot  \beta_L \left(1-L_{b1}\tim -L_{b2}\tim \right)^{ \beta_L-1} - \lambda_b\tim w_2\tim (1-\theta) + \lambda_{L2}\tim. \label{model-household-Lb2} 
\end{align}
For consumer goods, Eqs.~(\ref{model-firm-prod1}--\ref{model-firm-prod2}) have to be satisfied, guaranteeing that goods produced are identical to those consumed, invested, stored or delivered to the other sector.
Any ex-ante mismatch is compensated by adding constraint forces with factor $\tfrac{\partial Z_{P1}}{\partial C_{a1}} = \tfrac{\partial Z_{P2}}{\partial C_{a2}} = -1$ and Lagrangian multipliers $\lambda_{P1}$ and $\lambda_{P2}$ to the equation of motion.
The derivative of the budget constraint yields $\tfrac{\partial Z_a}{\partial C_{a1}} = p_1$, $\tfrac{\partial Z_a}{\partial C_{a2}} = p_2$.
The time evolution is given by:
\begin{align}
\dot C_{a1}\tim &= \mu_{aC1} \cdot \alpha_{C1} (C_{a1}\tim)^{\alpha_{C1}-1} (C_{a2}\tim)^{\alpha_{C2}} + \lambda_a\tim p_1\tim -\lambda_{P1}\tim, \label{model-household-Ca1} \\
\dot C_{a2}\tim &= \mu_{aC2} \cdot \alpha_{C2} (C_{a1}\tim)^{\alpha_{C1}} (C_{a2}\tim)^{\alpha_{C2}-1} + \lambda_a\tim p_2\tim -\lambda_{P2}\tim, \\
\dot C_{b1}\tim &= \mu_{bC1} \cdot \beta_{C1} (C_{b1}\tim)^{\beta_{C1}-1} (C_{b2}\tim)^{\beta_{C2}} +\lambda_b\tim p_1\tim -\lambda_{P1}\tim, \\
\dot C_{b2}\tim &= \mu_{bC2} \cdot \beta_{C2} (C_{b1}\tim)^{\beta_{C1}} (C_{b2}\tim)^{\beta_{C2}-1} +\lambda_b\tim p_2\tim -\lambda_{P2}\tim. \label{model-household-Cb2}
\end{align}
An extension to ``positional'' or ``conspicuous'' consumption will be modeled in Section \ref{sec_social} by adding a positive influence of household sector $b$ on consumption decisions by household sector $a$, and inversely.

The desired change in deposits held by households $\dot M_a$ and $\dot M_b$ reflects an intertemporal choice, but note that the bounded rational households cannot solve infinite optimization problems.
We assume that households value additional saving by the possible gain in leisure after a short period of time discounted by a factor $\rho_a$, at an average wage $(1-\theta)(w_1+w_2)/2$.
Combining this behavioral force with power factor $\mu_{aM}$ and the constraint force from the budget constraint with factor $\tfrac{\partial Z_a}{\partial M_a} = 1$ leads to:
\begin{align}
\dot M_a\tim &= \mu_{aM} (1+\alpha_r(r_M\tim - \rho_a)) \frac{ 2 \alpha_L \left(1-L_{a1}\tim -L_{a2}\tim \right)^{\alpha_L-1} }{(1-\theta)(w_1\tim + w_2\tim)}  + \lambda_a\tim, \label{model-household-MA} \\
\dot M_b\tim &= \mu_{bM} (1+ \beta_r(r_M\tim - \rho_b)) \frac{ 2  \beta_L \left(1-L_{b1}\tim -L_{b2}\tim \right)^{ \beta_L-1} }{(1-\theta)(w_1\tim + w_2\tim)} + \lambda_b\tim. \label{model-household-MB} 
\end{align}
The parameter $\alpha_r$ captures how strongly household sector $a$ considers this intertemporal choice.
For an alternative specification with a simple ``money in the utility function'' approach \citep{sidrauski_rational_1967}, see \citet{richters_modeling_2020}.

\subsubsection{Government}
\label{model-gov}

In this simple model, the government does not own assets or accumulates a stock of capital, but simply finances government consumption by tax income and credit.
The government has a disutility that grows with government debt:
\begin{align}
U_g\tim &= - (\gamma_D+\gamma_r r_g\tim) \frac{(D_g\tim)^2}{(p_1\tim+p_2\tim)/2} \Theta(D_g\tim),
\end{align}
with $\Theta$ the unit step function with $\Theta(D_g) = 0$ for $D_g \leq 0$ and $1$ otherwise.
Compared to disutility growing quadratically or as a power function \citep{cadenillas_explicit_2016}, this function is monotonously growing, avoiding negative debt to be considered adverse.
Further assets and roles for the government such as redistribution, market stabilization or provision of public goods may be implemented in the future.

The government tries to improve its utility, but note that the households' utility functions in Eqs.~(\ref{eq_utility_a}--\ref{eq_utility_b}) also contain a dependence on government values, which leads to an influence of the households on government decisions.
This shows how the GCD approach allows to formulate models with multiple, interacting economic forces with mutual or opposing interests.
Together with constraint forces proportional to $\frac{\partial Z_g}{\partial G_{g1}}$, $\frac{\partial Z_g}{\partial G_{g2}}$ and $\frac{\partial Z_g}{\partial D_g}$, this yields:
\begin{align}
\dot D_g\tim &= - \mu_{gD} \cdot (\gamma_D+\gamma_r r_g\tim) \frac{D_g\tim}{(p_1\tim+p_2\tim)/2} - \lambda_g\tim, \\
\dot G_{g1}\tim &= \mu_{aG1} \cdot \alpha_{G1} G_{g1}\tim^{\alpha_{G1}-1} + \lambda_g\tim p_1\tim -\lambda_{P1}\tim, \\
\dot G_{g2}\tim &= \mu_{bG2} \cdot \beta_{G2} G_{g2}\tim^{\beta_{G2}-1} + \lambda_g\tim p_2\tim -\lambda_{P2}\tim.
\end{align}
As in Eqs.~(\ref{model-household-Ca1}--\ref{model-household-Cb2}) for households, the constraint forces $\lambda_{P1}$ and $\lambda_{P2}$ correspond to ex-ante mismatches of supply and demand for goods.

As discussed above, the government sets taxation as proportional to labor income:
\begin{align}
0 &= \theta \left( w_1\tim L_{a1}\tim + w_2\tim L_{a2}\tim \right) - T_a\tim, \label{model-taxa} \\
0 &= \theta \left( w_1\tim L_{b1}\tim + w_2\tim L_{b2}\tim \right) - T_b\tim. \label{model-taxb}
\end{align}
These algebraic equations are equivalent to adding summands $(\theta ( w_1 L_{a1} + w_2 L_{a2} ) - T_a)^2$ and $(\theta ( w_1 L_{b1} + w_2 L_{b2} ) - T_b)^2$ to the utility function $U_g$, and this \emph{desire} being pursued with infinite power $\mu_{gT}$ \citep[see][]{richters_modeling_2020}.

\subsubsection{Firms in the production sectors}
\label{model-firms}

The firms in production sector $f1$ produce consumption goods for households $C_{a1} + C_{b1}$ and the government $G_{g1}$, change in inventories $\dot S_{f1}$, intermediate goods $A_{12}$ to be bought by sector $f2$, and gross investment consisting of replacement investment compensating depreciation $\delta_K K_{f1}$ and net investment $\dot K_{f1}$.
Firms' production $P_{f1}\tim$ is given by a Cobb-Douglas function with production inputs capital $K_{f1}$, labor $L_{f1}$ and intermediate goods $A_{21}$, see Eqs.~(\ref{model-firm-prod1}--\ref{model-firm-prod2}).
Different to many disequilibrium models that refute utility and substitutional production functions \citep{giraud_household_2019, godley_monetary_2012, berg_stock-flow_2015} referring to the Cambridge Capital Controversy \citep{harcourt_cambridge_1972}, we use a Cobb-Douglas production function.
But this does not imply that the profit maximizing input factors are always chosen instantaneously.
Instead, the adaptation process allows only for slow adaptation to profit opportunities.
Cobb-Douglas is chosen solely for tractability, replacing the production functions with CES or nested CES is straightforward, as is using a Leontief function.

The behavior of firms consists of an inventory and dividend policy, and the goal to increase their profits.
The firms in production sector $f1$ hold inventories $S_{f1}$ that act as a buffer stock against unexpected changes in demand.
From a modeling perspective, these buffer stocks are important as they avoid the system of equations to become stiff and unsolvable.
The targeted ratio $s_{f1}^\top$ of inventories to expected sales (gross investment plus sales to consumers, government, and sector $f2$) is constant.
The firms exert a force linearly increasing with the mismatch between targeted and actual inventories.
Similar to Eqs.~(\ref{model-household-Ca1}--\ref{model-household-Cb2}), a constraint force proportional to $\lambda_{P1}$ with factor $\tfrac{\partial Z_{P1}}{\partial S_{f1}} = - 1$ has to be added, assuming that every part of demand will be negatively affected if ex-ante demand is bigger than ex-ante supply, to guarantee ex-post consistency.
\begin{align}
\dot S_{f1}\tim &= \mu_{fS1}\left(s_{f1}^\top \left(C_{a1}\tim +C_{b1}\tim +G_{g1}\tim +A_{12}\tim +\delta_K K_{f1}\tim + \dot K_{f1}\tim \right) - S_{f1}\tim \right) -\lambda_{P1}\tim \nonumber \\
&= \mu_{fS1}\left(s_{f1}^\top \left( K_{f1}\tim^{\kappa_1} L_{f1}\tim^{l_1} A_{21}\tim^{1-\kappa_1-l_1} - \dot S_{f1}\tim \right) - S_{f1}\tim \right) -\lambda_{P1}\tim, \label{model-firm-Sf1dot} \\
\dot S_{f2}\tim 
&= \mu_{fS2}\left(s_{f2}^\top \left( K_{f2}\tim^{\kappa_2} L_{f2}\tim^{l_2} A_{12}\tim^{1-\kappa_2-l_2} - \dot S_{f2}\tim \right) - S_{f2}\tim \right) -\lambda_{P2}\tim. \label{model-firm-Sf2dot}
\end{align}
Using this specification, inventories can become negative, which can be understood as ``unfilled orders'' \citep[p.~907]{miron_seasonality_1988} or would have to be avoided using a strongly non-linear function.

Concerning the production factors, firms exert forces as gradients of their expected profits as utility functions $U_{f1}$ and $U_{f2}$.
Increasing production is costly not only because of direct inputs, but also because additional inventories according to Eqs.~(\ref{model-firm-Sf1dot}--\ref{model-firm-Sf2dot}) have to be financed by credit:
\begin{align}
\begin{split}
U_{f1} &= p_1 K_{f1}{}^{\kappa_1} L_{f1}{}^{l_1} A_{21}{}^{1-\kappa_1-l_1}-p_1 \delta_K K_{f1}-p_2 A_{21} -w_1 L_{f1} \\
& \qquad \ldots -r_{f1} p_1 \left( K_{f1} + s_{f1}^\top \left( K_{f1}{}^{\kappa_1} L_{f1}{}^{l_1} A_{21}{}^{1-\kappa_1-l_1} - \dot S_{f1} \right) \right),
\end{split}\\
\begin{split}
U_{f2} &= p_2 K_{f2}{}^{\kappa_2} L_{f2}{}^{l_2} A_{12}{}^{1-\kappa_2-l_2}-p_2 \delta_K K_{f2}-p_1 A_{12} -w_2 L_{f2} \\
& \qquad \ldots -r_{f2} p_2 \left( K_{f2} + s_{f2}^\top \left( K_{f2}{}^{\kappa_2} L_{f2}{}^{l_2} A_{12}{}^{1-\kappa_2-l_2} - \dot S_{f2} \right) \right).
\end{split}
\end{align}
Taking profits as basis for decision-making is similar to optimization approaches, but the difference is that firms do not jump directly to the point of highest profits by fully anticipating the reactions of households to changes in goods prices or wages.
Instead, firms try to increase their profits using a gradient seeking approach, only fully aware of the current marginal productivities and prices without any expectation about future sales.
The time evolution of the input factors consists of these profit driven forces and an additional constraint force with Lagrangian multiplier $\lambda_{P1}$ to satisfy the production equations (Eqs.~\ref{model-firm-prod1}--\ref{model-firm-prod2}) ex-post.

For capital, the economic force exerted by the firms is given by $\mu_{fK1} \frac{\partial U_{f1}}{\partial K_{f1}}$, while the prefactor for the Lagrangian multiplier is calculated as $\tfrac{\partial Z_{P1}}{\partial K_{f1}}$:
\begin{align}
\begin{split}
\dot K_{f1}\tim &= \mu_{fK1} \cdot p_1\tim \left( \left(1-r_{f1}\tim s_{f1}^\top\right) \kappa_1 K_{f1}\tim^{\kappa_1-1} L_{f1}\tim^{l_1} A_{21}\tim^{1-\kappa_1-l_1} -\delta_K - r_{f1}\tim \right) \\
& \qquad \ldots + \lambda_{P1}\tim \kappa_1 K_{f1}\tim^{\kappa_1-1} L_{f1}\tim^{l_1} A_{21}\tim^{1-\kappa_1-l_1},
\end{split} \label{model-firm-Kf1}  \\
\begin{split}
\dot K_{f2}\tim &= \mu_{fK2} \cdot p_2\tim \left( \left(1-r_{f2}\tim s_{f2}^\top\right) \kappa_2 K_{f2}\tim^{\kappa_2-1} L_{f2}\tim^{l_2} A_{12}\tim^{1-\kappa_2-l_2} -\delta_K - r_{f2}\tim \right) \\
& \qquad \ldots + \lambda_{P2}\tim \kappa_2 K_{f2}\tim^{\kappa_2-1} L_{f2}\tim^{l_2} A_{12}\tim^{1-\kappa_2-l_2}.
\end{split} \label{model-firm-Kf2} 
\end{align}
Note that total investment is given by $\dot K_{f1} + \delta_K K_{f1}$.

The time evolution of labor demand of firms contains an additional constraint force $- \lambda_{L1}$, added to guarantee consistency with labor supply by households according to Eqs.~(\ref{model-structure-L1}--\ref{model-structure-L2}).
\begin{align}
\begin{split}
\dot L_{f1}\tim &= \mu_{fL1}\left(p_1\tim \left(1-r_{f1}\tim s_{f1}^\top\right) l_1 K_{f1}\tim^{\kappa_1} L_{f1}\tim^{l_1-1} A_{21}\tim^{1-\kappa_1-l_1} - w_1\tim \right) \\
& \qquad \ldots + \lambda_{P1}\tim l_1 K_{f1}\tim^{\kappa_1} L_{f1}\tim^{l_1-1} A_{21}\tim^{1-\kappa_1-l_1} - \lambda_{L1}\tim,
\end{split}  \label{model-firm-Lf1} \\
\begin{split}
\dot L_{f2}\tim &= \mu_{fL2}\left(p_2\tim \left(1-r_{f2}\tim s_{f2}^\top\right) l_2 K_{f2}\tim^{\kappa_2} L_{f2}\tim^{l_2-1} A_{12}\tim^{1-\kappa_2-l_2} - w_2\tim \right) \\
& \qquad \ldots + \lambda_{P2}\tim l_2 K_{f2}\tim^{\kappa_2} L_{f2}\tim^{l_2-1} A_{12}\tim^{1-\kappa_2-l_2} - \lambda_{L2}\tim.
\end{split}  \label{model-firm-Lf2} 
\end{align}
If labor is cheap compared to its contribution to production, the labor input is increased, but not instantaneously, and the constraint forces can lead to deviations from this plan.

For intermediate goods $A_{21}$ produced by sector $f2$ and used by sector $f1$, the time evolution contains an additional term $\lambda_{P2}$ with factor $\frac{\partial Z_{P2}}{\partial A_{21}} = -1$, because sector $f1$ is affected if there is insufficient production in sector $f2$:
\begin{align}
\begin{split}
\dot A_{21}\tim &= \mu_{fA1} \left(p_1\tim \left(1- r_{f1}\tim s_{f1}^\top\right)\left(1-\kappa_1-l_1\right)K_{f1}\tim^{\kappa_1} L_{f1}\tim^{l_1} A_{21}\tim^{-\kappa_1-l_1} - p_2\tim \right) \\
& \qquad \ldots + \lambda_{P1}\tim \left(1-\kappa_1-l_1\right) K_{f1}\tim^{\kappa_1} L_{f1}\tim^{l_1} A_{21}\tim^{1-\kappa_1-l_1} -\lambda_{P2}\tim,
\end{split}\\
\begin{split}
\dot A_{12}\tim &= \mu_{fA2} \left(p_2\tim \left(1- r_{f2}\tim s_{f2}^\top\right)\left(1-\kappa_2-l_2\right)K_{f2}\tim^{\kappa_2} L_{f2}\tim^{l_2} A_{12}\tim^{-\kappa_2-l_2} - p_1\tim \right) \\
& \qquad \ldots + \lambda_{P2}\tim \left(1-\kappa_2-l_2\right) K_{f2}\tim^{\kappa_2} L_{f2}\tim^{l_2} A_{12}\tim^{1-\kappa_2-l_2} -\lambda_{P1}\tim.
\end{split}
\end{align}

The dividend policy is such that distributed profits $\pi_{f1}$ and $\pi_{f2}$ are total production minus input costs, which implies that changes in value of existing capital are not distributed:
\begin{align}
0 &= p_1\tim K_{f1}\tim^{\kappa_1} L_{f1}\tim^{l_1} A_{21}\tim^{1-\kappa_1-l_1} - p_1\tim \delta_K K_{f1}\tim - p_2\tim A_{21}\tim -w_1\tim L_{f1}\tim -r_{f1}\tim D_{f1}\tim - \pi_{f1}\tim, \label{model-firm-pif1} 
\\
0 &= p_2\tim K_{f2}\tim^{\kappa_2} L_{f2}\tim^{l_2} A_{12}\tim^{1-\kappa_2-l_2} - p_2\tim \delta_K K_{f2}\tim - p_1\tim A_{12}\tim -w_2\tim L_{f2}\tim -r_{f2}\tim D_{f2}\tim - \pi_{f2}\tim. \label{model-firm-pif2}
\end{align}
This is an example of a behavioral equation implemented as an algebraic equation, implying that $\pi_{f1}$ and $\pi_{f2}$ are \emph{not} influenced by constraint forces.
Alternatively, a principal--agent dilemma could be modeled by incorporating individual forces of shareholders trying to increase dividends while the management may favor retained earnings \citep{la_porta_agency_2000}.
Using the accounting and budget constraints in Eqs.~(\ref{balance-sheet-f1}--\ref{balance-sheet-f2}, \ref{model-budget-f1}--\ref{model-budget-f2}), the time evolution of credit and equity can be calculated to be:
\begin{align}
\dot D_{f1}\tim = p_1\tim (\dot K_{f1}\tim + \dot S_{f1}\tim), \label{model-firm-dotD1} \\
\dot D_{f2}\tim = p_2\tim (\dot K_{f2}\tim + \dot S_{f2}\tim), \label{model-firm-dotD2} \\
\dot E_{f1}\tim = \dot p_1\tim (K_{f1}\tim + S_{f1}\tim), \\
\dot E_{f2}\tim = \dot p_2\tim (K_{f2}\tim + S_{f2}\tim).
\end{align}
Thus new investment is financed by credit, while changes in value of existing capital changes the equity of firms:
$D_{f1}$, $D_{f2}$, $E_{f1}$ and $E_{f1}$ adapt to satisfy the constraints and no Lagrangian multipliers $\lambda_{f1}$ and $\lambda_{f2}$ are necessary to guarantee consistency.
Using these assumptions, there is no feedback from net worth on costs or volumes of external finance.

\subsubsection{Banking sector}

The balance sheet and budget constraint of the banking sector are Eqs.~(\ref{balance-sheet-bank}, \ref{model-budget-bank}).
Banks are rather passive actors in this model:
They lend money ``on demand'' at the current interest rates $r_{f1}$, $r_{f2}$ and $r_g$ to firms and the government in line with the concept of endogenous money creation \citep[see][]{wray_money_1990, gross_money_2019}. They pay interest $r_M M_a$ and $r_M M_b$ to households and distribute all their profits $\pi_{bank}$ to the two household sectors, here implemented as an algebraic equation:
\begin{align}
0 &= r_{f1}\tim D_{f1}\tim + r_{f2}\tim D_{f2}\tim + r_g\tim D_g\tim - r_M\tim \left(M_a\tim +M_b\tim \right) - \pi_{bank}\tim. \label{model-pibank}
\end{align}
A richer behavioral model of banks that includes credit rationing or agency costs may be integrated in the future.

\subsubsection{Price development}

The prices react to ex-ante mismatches between supply and demand. 
If the agents' plans would increase demand stronger than supply, the firms realize that they are unable to change their inventories as desired, which is the case if the Lagrangian multiplier $\lambda_{P1} > 0$.
This leads to rationing for all buyers of this good, thus households, government, the other sector, but also for capital investment in the own sector.
With the constraint forces based on Eqs.~(\ref{eq_economics_3}--\ref{eq_economics_4}), the rationing is symmetric in the sense that every flow of goods is affected in the same way.
As these constraint forces in economics cannot be derived from laws of nature, they can be varied to allow for asymmetric constraints.

Parallel to rationing, sector $f1$ slowly increases the price $p_1$ with a linear reaction function to differences between the ex-ante values of supply and demand: 
\begin{align}
\dot p_1\tim &= \mu_{p_1} \lambda_{P1}\tim, \label{model-p1dot} \\
\dot p_2\tim &= \mu_{p_2} \lambda_{P2}\tim.
\end{align}
Similarly, the wages react on a mismatch between ex-ante supply and demand for labor:
\begin{align}
\dot w_1\tim &= \mu_w \lambda_{L1}\tim, \label{model-w1dot} \\
\dot w_2\tim &= \mu_w \lambda_{L2}\tim. \label{model-w2dot}
\end{align}
The interest rates are adapted by the central bank following a simple inflation targeting rule. If the average price change is above a target $\rho^\top$, interest rates are increased:
\begin{align}
\dot r_g &= \dot r_{f1} = \dot r_{f2} = \dot r_{M} = \mu_r \left(\frac{\dot p_1 / p_1 + \dot p_2 / p_2}{2} - \rho^\top \right).
\end{align}
An increased nominal interest rate has a negative impact on firms' investment (Eqs.~\ref{model-firm-Lf1}--\ref{model-firm-Lf2}) and a positive impact on households' saving decision (Eqs.~\ref{model-household-MA}--\ref{model-household-MB}), which leads to a reduction of demand and inflation, see the column designated with $r_g$ in Figure~\ref{fig:Jacobian-heatmap}.
This monetary policy rule can be extended to include further policy mandates of central banks, such as reacting to deviations from a targeted level of economic activity or employment \citep{taylor_discretion_1993, taylor_simple_2021}.

All the parameters $\mu$ and the choice of the factors influencing price development reflect assumptions about power relations and adaptation speeds within the economy.

\section{Model analysis}
\label{model-analysis}

\subsection{Time evolution and stationary states}
\label{model-time-evolution}

The initial conditions have to satisfy the six balance sheet constraints (Eqs.~\ref{balance-sheet-f1}--\ref{balance-sheet-g}), the two labor constraints (Eqs.~\ref{model-structure-L1}--\ref{model-structure-L2}) and the five algebraic equations for taxation (Eqs.~\ref{model-taxa}--\ref{model-taxb}) and profit distribution (Eqs.~\ref{model-firm-pif1}--\ref{model-firm-pif2}, \ref{model-pibank}).
No further equilibrium conditions are presupposed.

\begin{figure*}[tp]
\hfil\includegraphics[width=\textwidth]{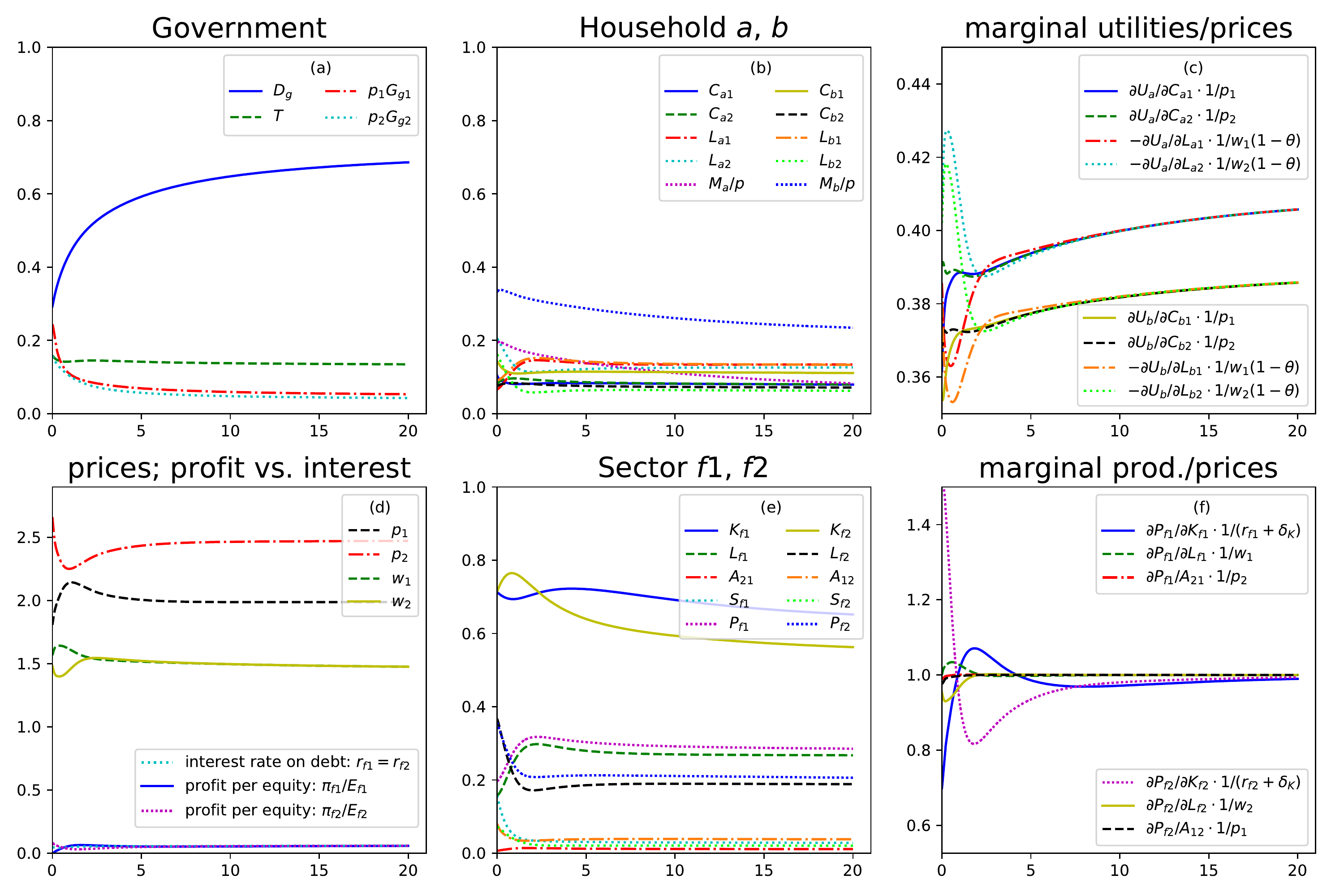}\hfil 
\caption{\label{fig:time-evolution-1}Plots (a), (b), (d) and (e) show the time evolution of the variables for different sectors. Plot (c) shows that for the two household sectors, the marginal utilities for consumption and leisure, divided by their respective price, equalize over time. The budget equation constrains their choices, and the gradient climbing approach converges to the highest reachable level of utility.
The same can be stated for plot (f) concerning the marginal input productivities for capital, labor and intermediate goods, divided by their price. Plot (d) shows that in equilibrium, profit paid per unit of equity is identical to the interest rate paid on credit.
}
\end{figure*}

As an example, Figure~\ref{fig:time-evolution-1} shows the time evolution for the initial conditions, power factors and further parameters summarized in Appendix~\ref{sec_initial_conditions}.
At $t=0$, plot~(c) shows that for household sector $b$, the marginal utility of leisure divided by the wage $\frac{\partial U_b}{L_{b1}} \frac{1}{(1-\theta) w_1}$ is higher than the marginal utility of consuming good 1 divided by its price $\frac{\partial U_b}{C_{b1}} \frac{1}{p_1}$, and for good $C_{b2}$ this value is even lower.
Therefore, the forces of household sector $b$ try to push the economy towards reducing work and consuming less, particularly of good 2.
For household sector $a$, $\frac{\partial U_a}{C_{a2}} \frac{1}{p_2}$ is higher than $\frac{\partial U_a}{C_{a1}} \frac{1}{p_1}$, thus his forces try increase consumption of good $C_{a2}$ compared to $C_{a1}$.
Plot~(f) compares the marginal productivities of inputs divided by their respective price. At $t=0$, the marginal productivity of capital in sector $f1$ is lower than the interest rate.
This is the reason why profits per equity $\pi_{f1}/E_{f1}$ in sector $f1$ are very low, see plot~(d).
To improve profits, this sector exerts forces to reduce $K_{f1}$.
Sector $f2$ is in the opposite situation, trying to increase $K_{f2}$.

The time evolution created by these ex-ante forces would not satisfy the constraints.
For example, the changes in demand and supply for good 1 create a tendency of excess demand ($\lambda_{P1} > 0$).
The corresponding constraint forces influence the dynamics such that the constraints are satisfied ex-post.
Additionally, the price $p_1$ increases according to Eq.~\eqref{model-p1dot}, while there is a tendency for excess supply for good 2, leading to a negative slope of $p_2$.
The adjustment processes for quantities and prices ultimately converge to a stationary state whose properties can be calculated analytically.

\subsection{Properties of stationary states}
\label{model-stationary}

The fixed point of a model is one with vanishing time derivatives in every variable.
To derive the conditions for the stationary state, assume that every power factor is positive.
From the price development (Eqs.~\ref{model-p1dot}--\ref{model-w2dot}), it follows that $\lambda_{P1} = \lambda_{P2} = 0$ and $\lambda_{L1} = \lambda_{L2} = 0$, thus in the stationary state, there is no mismatch between supply and demand for labor and goods.

For sector $f1$, the following conditions hold:
\begin{align}
0 &= p_1( C_{a1}+ C_{b1} +G_{g1}+A_{12})-A_{21}p_2-w_1L_{f1}-r_{f1}D_{f1}-\pi_{f1}, \label{stability-analysis1} \\
0 &= s_{f1}^\top \left(C_{a1}+C_{b1}+G_{g1}+A_{12}+\delta_K K_{f1}\right) - S_{f1}, \\
0 &= K_{f1}^{\kappa_1} L_{f1}^{l_1} A_{21}^{1-\kappa_1-l_1} -\delta_K K_{f1}-C_{a1}- C_{b1} -G_{g1}-A_{12}, \label{stability-analysis2} \\
0 &= (1-r_{f1}s_{f1}^\top) \kappa_1 K_{f1}^{\kappa_1-1} L_{f1}^{l_1} A_{21}^{1-\kappa_1-l_1} -\delta_K - r_{f1}, \\
0 &= p_1(1-r_{f1}s_{f1}^\top) l_1 K_{f1}^{\kappa_1} L_{f1}^{l_1-1} A_{21}^{1-\kappa_1-l_1} - w_1, \\
0 &= p_1(1- r_{f1}s_{f1}^\top)(1-\kappa_1-l_1)K_{f1}^{\kappa_1} L_{f1}^{l_1} A_{21}^{-\kappa_1-l_1} - p_2. \label{stability-analysis6} 
\end{align}
With an inventory target of $s_{f1}^\top = 0$, the factor share is identical to the output elasticity, the exponent of the production factor in the Cobb-Douglas function, as in neoclassical competitive equilibrium.
With $s_{f1}^\top > 0$, a part of total income goes to interest payments related to inventory holding that do not contribute to increased production.

Using the definition (valid because $\dot K_{f1}= \dot S_{f1}= 0$)
\begin{align}
P_{f1}&= K_{f1}^{\kappa_1} L_{f1}^{l_1} A_{21}^{1-\kappa_1-l_1} = \delta_K K_{f1}+ C_{a1}+ C_{b1}+ G_{g1}+ A_{12},
\end{align}
Eqs.~(\ref{stability-analysis2}--\ref{stability-analysis6}) can be simplified to:
\begin{align}
p_1S_{f1}&= p_1s_{f1}^\top P_{f1}, \\
(r_{f1}+\delta_K) p_1K_{f1}&= p_1  (1-r_{f1}s_{f1}^\top) \kappa_1          P_{f1}, \\
p_2A_{21}                &= p_1  (1-r_{f1}s_{f1}^\top) (1-\kappa_1-l_1)  P_{f1}, \\
w_1L_{f1}                 &= p_1  (1-r_{f1}s_{f1}^\top) l_1               P_{f1}.
\end{align}
The credit is given by:
\begin{align}
 D_{f1}&= p_1(K_{f1}+ s_{f1}^\top P_{f1}) - E_{f1}.
\end{align}
Substituting these results in the definition of profit $\pi_{f1}$ in Eq.~\eqref{model-firm-pif1} yields (see Appendix~\ref{appendix-pif1}):
\begin{align}
\pi_{f1}&= p_1\left( C_{a1}+C_{b1}+G_{g1}+A_{12}\right) - A_{21}p_2-w_1L_{f1}-r_{f1}D_{f1}= r_{f1}E_{f1}.
\end{align}
This derivation shows that in the stationary state, the profits are a compensation for equity capital $E_{f1}$, and both equity capital and credit have the same rate of return, see Figure~\ref{fig:time-evolution-1}(d).
This corresponds to the first theorem by \citet{modigliani_cost_1958}, assuming that no financial frictions and no difference in riskiness exists.

For household sector $a$, we assume that $\mu_{aC1} = \mu_{aC2} = \mu_{aM} = \mu_{aL1} = \mu_{aL2}$, implying that households have the same power to influence all their variables. The following conditions hold:
\begin{align}
0 &= p_1C_{a1}+ p_2C_{a2}- (1-\theta) \left( w_1L_{a1}+ w_2L_{a2}\right) - r_MM_a- e_a (\pi_{f1}+ \pi_{f2}+\pi_{bank}), \\
0 &= \alpha_{C1} \left(C_{a1}\right)^{\alpha_{C1}-1} \left(C_{a2}\right)^{\alpha_{C2}} +\lambda_ap_1, \\
0 &= \left(C_{a1}\right)^{\alpha_{C1}} \alpha_{C2} \left(C_{a2}\right)^{\alpha_{C2}-1} +\lambda_ap_2, \\
0 &= -\alpha_L \left(1-L_{a1}-L_{a2}\right)^{\alpha_L-1} - \lambda_aw_1(1-\theta), \\
0 &= -\alpha_L \left(1-L_{a1}-L_{a2}\right)^{\alpha_L-1} - \lambda_aw_2(1-\theta), \\
0 &= r_M - \rho_a.
\end{align}
The equations imply that total income from wages and capital is equal to taxes and consumption, and that the wages in both sectors have to be identical.
Canceling $\lambda_a$ from all the equations yields the first order conditions for consumers in general equilibrium models: 
\begin{align}
- \frac{\partial U_a / \partial L_{a1}}{(1-\theta)w_1} = - \frac{\partial U_a / \partial L_{a2}}{(1-\theta)w_2} = \frac{\partial U_a / \partial C_{a1}}{p_1} = \frac{\partial U_a / \partial C_{a2}}{p_2}.
\end{align}
The ratio of prices equals the ratio of marginal utilities, thus the utility from the last monetary unit spent on each good must be the same and identical to the disutility of increasing working time divided by the wage after tax $(1-\theta)w_1$.
In equilibrium, the interest rate on deposits $r_M$ equals the rate of time preference $\rho_a$.
Note that this stationary state can be reached if and only if $\rho_a = \rho_b$, as $r_M$ cannot converge to two distinct values simultaneously.
If $\rho_a > \rho_b$, household sector $a$ accumulates debt to finance consumption, as a no-ponzi condition is missing in this model.
One way to relax this condition in the future would be to let the bank charge heterogeneous interest rates, depending on the debt-income ratio, which would allow the interest rate on credit for household sector $a$ to rise to $\rho_a$.

The total income distributed from sector $f1$ to household sectors $a$ and $b$ before taxation is given by (see Appendix~\ref{appendix-pif1}):
\begin{align}
\pi_{f1}+ r_{f1}D_{f1}+ w_1L_{f1}= p_1P_{f1}- p_2A_{21}- \delta_K p_1K_{f1}.
\end{align}
Total income is equal to production minus intermediate purchases minus depreciation.

Overall, with respect to households and firms, the stationary state satisfies all the condition usually presupposed in static neoclassical general equilibrium models.
This result is independent on the power factors.
In this specification of the model, economic power influences the adaptation processes, but not the equilibrium reached.

For the government, the equations in the stationary state are, assuming $\mu_{aG1} = \mu_{bG2} = \mu_{gD}$:
\begin{align}
0 &= \alpha_{G1} G_{g1}^{\alpha_{G1}-1} + \lambda_gp_1, \\
0 &= \beta_{G_{g2}} G_{g2}^{\beta_{G2}-1} + \lambda_gp_2, \\
0 &= -2 (\gamma_D +\gamma_r r_g) D_g/(p_1+p_2) - \lambda_g, \\
0 &= r_gD_g+ p_1G_{g1}+ p_2G_{g2}- T_a- T_b, \\
0 &= \theta p_1(1-r_{f1}s_{f1}^\top) l_1 P_{f1}+ \theta p_2(1-r_{f2}s_{f2}^\top) l_2 P_{f2}- T_a- T_b.
\end{align}
In the stationary state, tax income covers government expenditures and interest payments on government credit $D_g$. The economic force applied by the households is counterbalanced by politicians desire to limit government debt.

\subsection{Local and global stability}

\begin{figure*}[tp]
\centering
\includegraphics[height=0.8\textwidth]{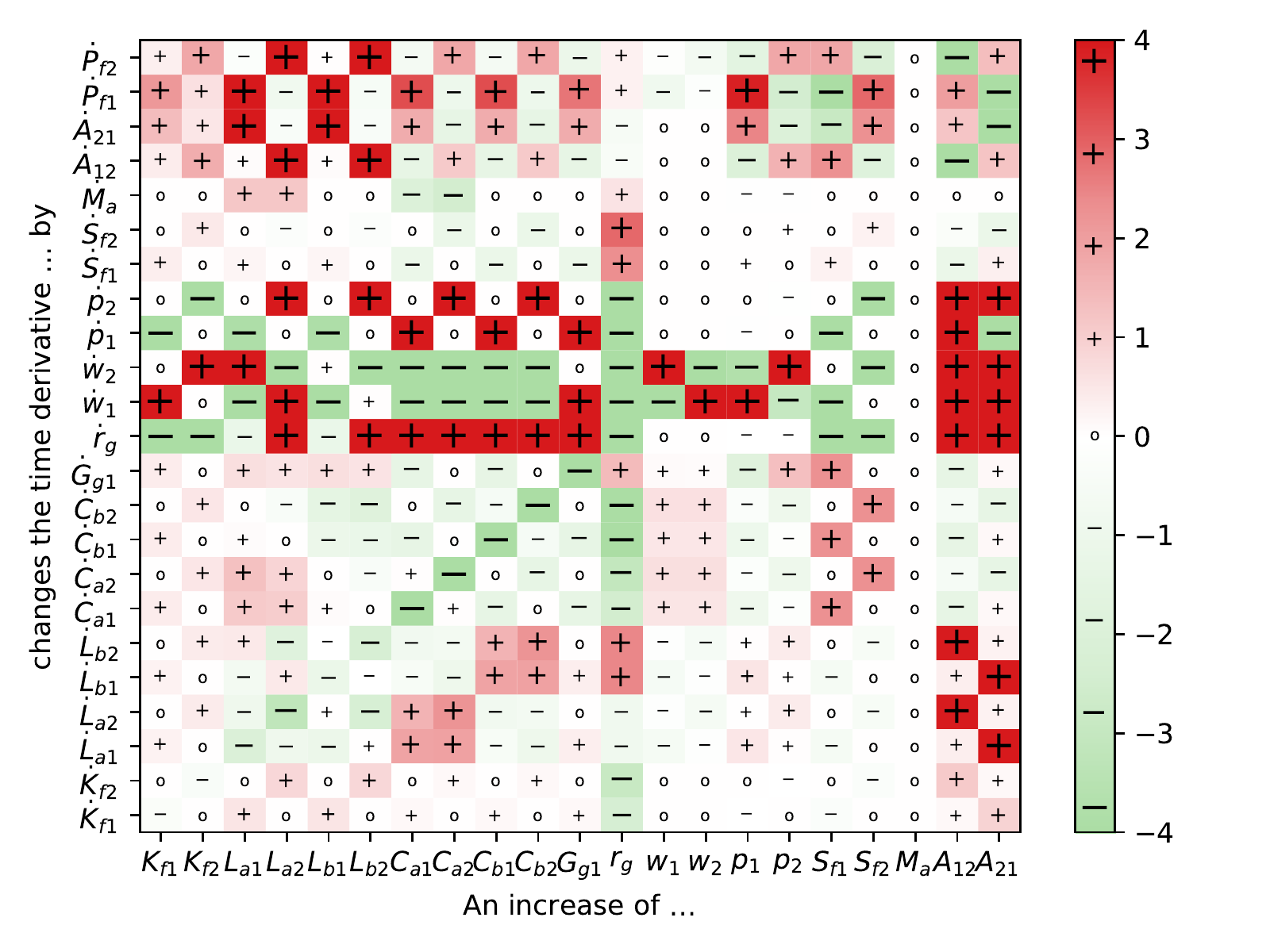}
\caption{\label{fig:Jacobian-heatmap}This selection of matrix entries of the Jacobian $J_T$ in Eq.~\eqref{model:jacobian} illustrates the impact of a small increase in one of the variables on the time evolution of the others. The reactions of the time derivatives to deviations from the equilibrium can be extracted from the diagram. \newline
For example, the penultimate column implies that an increase in intermediate trade $A_{12}$ from sector $f1$ to $f2$ leads to a reduction of the inventory stock ($\dot S_{f1} < 0$), which leads to an increase in price $\dot p_1 > 0$, increasing inputs $\dot L_{a1}, \dot L_{b1}, \dot K_{f1}, \dot A_{21}$ and a negative time evolution of the other sales $\dot C_{a1}, \dot C_{b1}$ and $\dot G_{g1}$.
The other inputs $L_{a2}, L_{b2}, K_{f2}$ grow, because the additional input $A_{12}$ increases their marginal productivities.
The additional demand for labor and capital leads to increasing wages and interest rates.
These rising costs together with lowered prices $p_2$ will reverse this development in the following and push the economy back to equilibrium.}
\end{figure*}

\begin{figure*}[tp]
\centering
\includegraphics[height=0.8\textwidth]{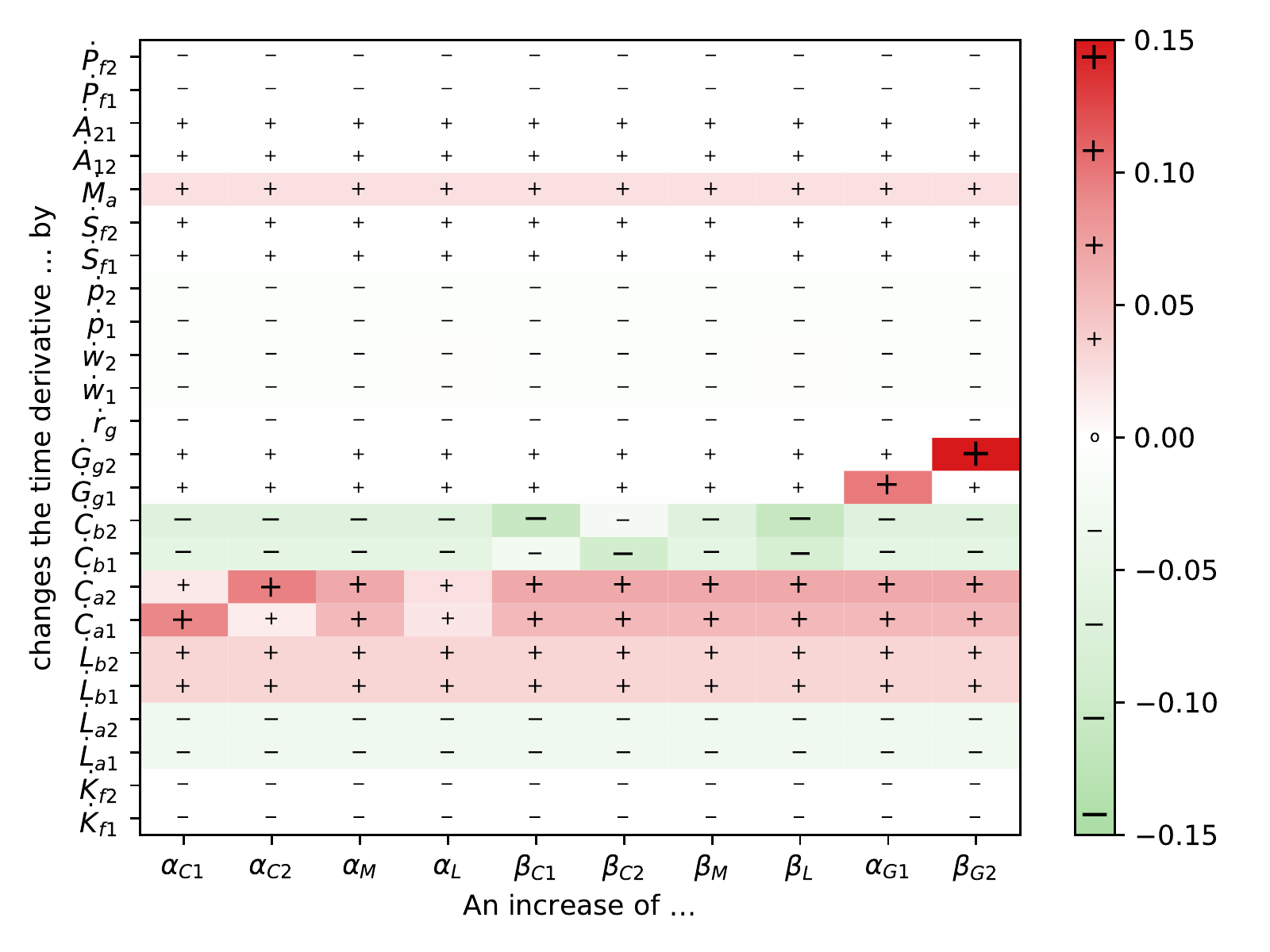}
\caption{\label{fig:Parameter-heatmap}These matrix entries depict the impact of a small increase in one of the parameters on the time evolution of the variables. \newline
For example, if the exponent $\alpha_{C1}$ of household consumption $C_{a1}$ is increased, this leads to a shift of consumption by household sector $a$ from production sector $f2$ to $f1$.
If $\beta_{G2}$ is increased and therefore the impact of household sector $b$ on government expenditure $G_2$, this variable increases.}
\end{figure*}

The differential-algebraic equation framework poses a challenge for the local stability analysis. Because of the constraints, the variables cannot be varied independently:
A change in working hours necessarily implies a change in production, inventories, wage income, saving etc.
The six balance sheet constraints (Eqs.~\ref{balance-sheet-f1}--\ref{balance-sheet-g}), the two labor constraints (Eqs.~\ref{model-structure-L1}--\ref{model-structure-L2}) and the five algebraic equations for taxation (Eqs.~\ref{model-taxa}--\ref{model-taxb}) and profit distribution (Eqs.~\ref{model-firm-pif1}--\ref{model-firm-pif2}, \ref{model-pibank}) have to be guaranteed even after the shock.
Additionally, interest rates have to march in lockstep, $\dot r_g = \dot r_{f1} = \dot r_{f2} = \dot r_{M}$, and $E_{bank}=0$.
These 17 restrictions have to be fulfilled, and we chose $T_a$, $T_b$, $L_{f1}$, $L_{f2}$, $r_{f1}$, $r_{f2}$, $r_M$, $E_{f1}$, $E_{f2}$, $E_{bank}$, $V_a$, $V_b$, $D_g$, $V_g$, $\pi_1$, $\pi_2$ and $\pi_{bank}$ to be determined by constraints, while the remaining 25 values are varied: $x = \{K_{f1}$, $K_{f2}$, $L_{a1}$, $L_{a2}$, $L_{b1}$, $L_{b2}$, $C_{a1}$, $C_{a2}$, $C_{b1}$, $C_{b2}$, $G_{g1}$, $G_{g2}$, $r_g$, $w_1$, $w_2$, $p_1$, $p_2$, $S_{f1}$, $S_{f2}$, $M_{a}$, $M_{b}$, $D_{f1}$, $D_{f2}$, $A_{12}$, $A_{21}$\}.
This choice is a bit arbitrary: For example, looking at the labor constraint $0 = L_{f1} - L_{a1} - L_{b1}$, two of the variables can be varied, while the third is dependent on the two others.
If you want to increase $L_{a1}$ while keeping $L_{f1}$ fixed, then this is realized by simultaneously decreasing $L_{b1}$.
This way, every variation consistent with the constraints can be obtained.
The production constraints (Eqs.~\ref{model-firm-prod1}--\ref{model-firm-prod2}) are not problematic because the change in inventories $\dot S_{f1}$, $\dot S_{f2}$ can absorb the shock.
The time evolution $\dot{x} = T(x)$ around the equilibrium $x_{eq}$ can be linearized with the $(25 \times 25)$ Jacobian matrix $J_T$ of all the first-order partial derivatives:
\begin{align}
J_T(x_{eq}) := \left(\frac{\partial T_i}{\partial x_j}(x_{eq})\right)_{i,j=1,\ldots,n} =  \begin{pmatrix}
\frac{\partial T_1}{\partial x_1}(x_{eq}) & \frac{\partial T_1}{\partial x_2}(x_{eq}) & \ldots & \frac{\partial T_1}{\partial x_n}(x_{eq}) \\
\vdots & \vdots & \ddots & \vdots \\
\frac{\partial T_n}{\partial x_1}(x_{eq}) & \frac{\partial T_n}{\partial x_2}(x_{eq}) & \ldots & \frac{\partial T_n}{\partial x_n} (x_{eq}) \label{model:jacobian}
\end{pmatrix}.
\end{align}
The Jacobian matrix contains the reaction of the economy to a shock in equilibrium, as illustrated in Figure~\ref{fig:Jacobian-heatmap} for a shock to the variables and Figure~\ref{fig:Parameter-heatmap} for changes in some parameter values.
For other aspects such as the exponents in the production function, the change in the stationary state is comparable to what is known from General Equilibrium models.

The relevant quantities for the first order stability of the stationary state are the eigenvalues of the Jacobian $J_T$.
An analytical calculation shows that $0$ is a double eigenvalue, thus $J_T v_i = 0$ with $v_i$ being two corresponding linearly independent eigenvectors.
The first eigenvector corresponds to an increase of $L_{a1}$ and $L_{b2}$ by $\Delta L$, while $L_{a2}$ and $L_{b1}$ are reduced by the same amount:
Household sector $a$ works longer in sector $f1$, but shorter in sector $f2$, and household sector $b$ inversely.
The aggregated variables $L_a$, $L_b$, $L_{f1}$ and $L_{f2}$ remain unchanged.
The second eigenvector corresponds to an increase of $D_{f1}$ and $E_{f2}$ by $\Delta D$, accompanied by a decrease of $D_{f2}$ and $E_{f1}$ by the same amount.
Sector $f1$ is now financed to a larger share by credit instead of equity, while it is the inverse for sector $f2$.
Correspondingly, interest payments by sector $f1$ are increased while distributed profits are decreased and inversely for sector $f2$, keeping total equity and total firms' debt unchanged.
In both cases, the stationary state is not unique but path dependent in some microscopic variables, but sectoral production, allocation, distribution and consumption remain unchanged.

\begin{figure*}[tp]
\centering \includegraphics[width=\textwidth]{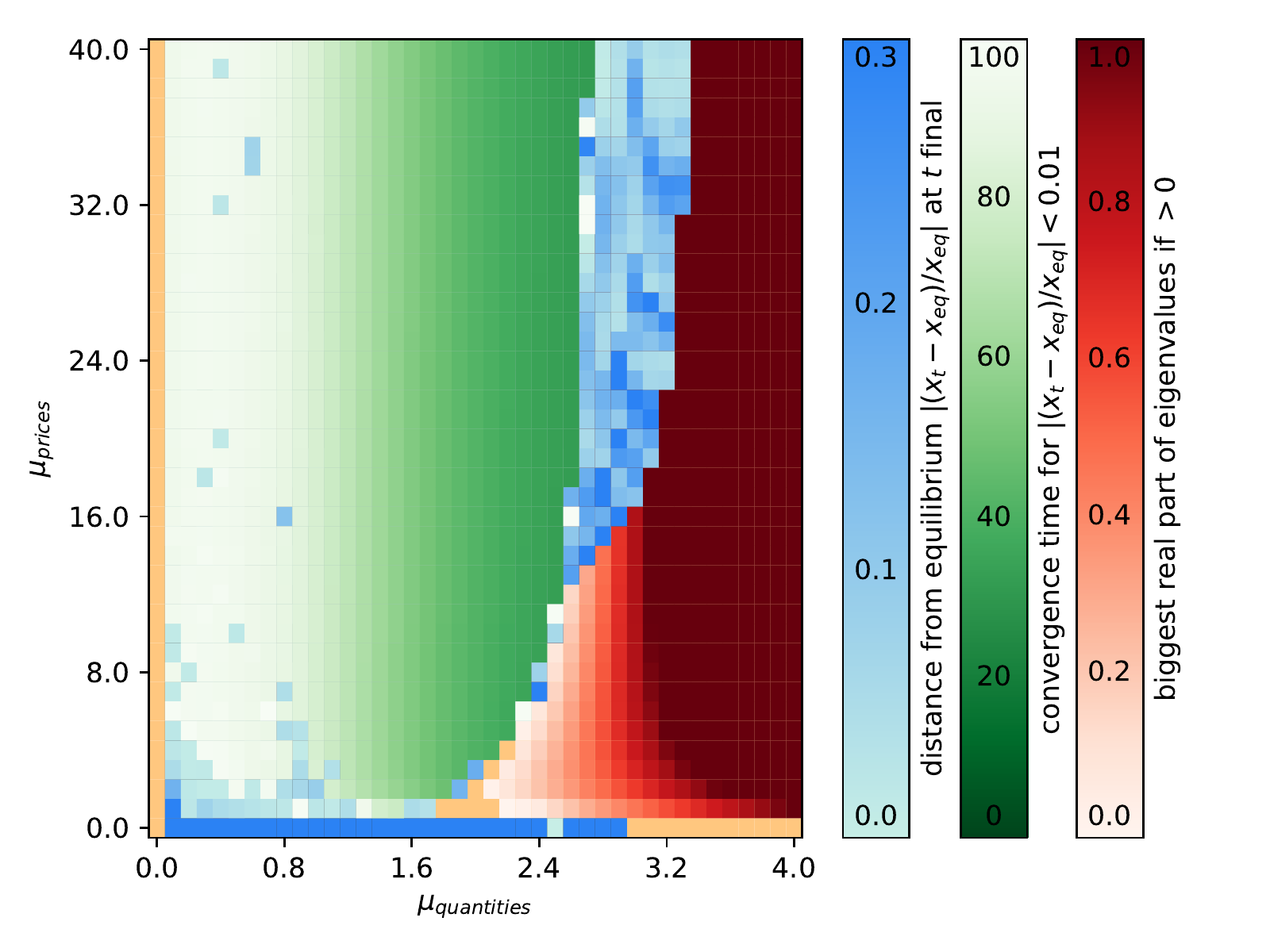}
\caption{\label{fig:stability-analysis-1}Stability analysis of the system, depending on the scaling factors $\mu_{prices}$ and $\mu_{quantities}$. The red color is used for combinations in which the biggest real part of the eigenvalues of the Jacobian $J_T$ at the stationary state is above zero. Therefore, the system is locally unstable, the model shows explosive behavior and the numerical solver aborts. The orange color indicates the area where the real parts of all the eigenvalues are $\leq 0$, but the numerical solver aborts nevertheless. This part of the system is locally stable, but shows no convergence for the initial conditions. The green part converges to the numerically determined equilibrium. The greener the color, the faster the convergence until $|(x_t-x_{eq})/x_{eq}| < 0.01$. In the blue part, the system did not converge to a stationary state at $t = 100$. The difference $|(x_t-x_{eq})/x_{eq}|$ at $t = 100$ is indicated by the blue color. For $\mu_{quantities} = 0$, the model does not converge because the individual influences on quantities are negligible. For $\mu_{prices}=0$ and $\mu_{quantities} > 0$, the model does not converge to a stationary state as the coordinating influence of price adaptation is missing.}
\end{figure*}

The other eigenvalues depend on the parameters, particularly the power factors $\mu$, as revealed by the stability analysis in Figure~\ref{fig:stability-analysis-1}.
Starting from the parameters in Section \ref{model-time-evolution}, each power factor related to quantities (such as $\mu_{aC1}$, $\mu_{fK1}$, $\mu_{gD}$, \ldots) is multiplied by a common factor $\mu_{quantities}$, while power factors related to prices (such as $\mu_w$, $\mu_{p_1}$, \ldots) are multiplied by a factor $\mu_{prices}$.
In the red part on the right, the biggest real part of the eigenvalues is bigger than zero, implying local instability.
For $\mu_{quantities}$ big, the quantities react so strongly for example on profit opportunities that the oscillations of the system become unstable.
The stationary state in the small orange part is locally stable, but the time evolution does abort because either capital or intermediate goods drop to zero during the adaptation process.
If $\mu_{quantities} = 0$, the numerical solver aborts because no market forces prevent capital or labor from taking negative values, leading to an undefined value of the production function.
In the green part, the time evolution converges to the equilibrium derived in Section \ref{model-stationary}.
In the blue part, the system did not converge to a stationary state at $t = 100$, showing oscillations similar to Figure~\ref{fig:time-evolution-2}.
Varying the power factors related to quantities independently revealed that the crucial factors responsible for the instability are the power factors $\mu_{fK1}$ and $\mu_{fK2}$ that govern the investment reaction of firms to profit opportunities.
A similar result is obtained by \citet[p.~237]{godley_monetary_2012} arguing that models become unstable if investment reacts strongly to changes.

An analytical global stability analysis of the 25 equation model is not possible.
We therefore randomized the initial conditions, solved the model numerically and compared the stationary states.
Independent of the power factors, the solver sometimes aborts quickly.
An analysis of these cases revealed that either fixed capital or intermediate goods were pushed down to zero by the constraint forces, if the model was initialized at a situation with strong excess demand and therefore rationing.
The price adaptation processes were then not fast enough to stabilize the system.
If the economy survived this transient situation of economic imbalances and was within the suitable range of power parameters according to Figure \ref{fig:stability-analysis-1}, all models converge to the same equilibrium with respect to production decisions (input factors, total production).
The financing decision varies depending on initial conditions, because in equilibrium, there is no difference between debt and equity financing.
Also, as households are indifferent between working in sector $f1$ or $f2$ once wages have equilibrated, the stationary states differ in which household works in which production sector, but not with respect to total work of each household or in each sector.

Outside the stable region of Figure \ref{fig:stability-analysis-1}, the reaction functions of the economic actors and the price adaptation cannot guarantee global stability.
If quantity adjustments are fast, the model becomes unstable, because the bounded rational firms do not anticipate the reactions of the other market participants to their change in production.
Instead, they react on supply--demand mismatches by adapting production and prices.
If this reaction is very strong, it can lead to growing inventory oscillations as pioneered by \citet{metzler_nature_1941}.
Faster price adaptation can sometimes improve local stability.
Different from the way frictions are commonly discussed in economic models as slowing down convergence to the equilibrium, very fast adaptations of quantities make the equilibrium unattainable.
In this model, intermediate adaptation speeds lead to the fastest convergence to equilibrium.

\subsection{Social interaction, fiscal and monetary policy}
\label{sec_social}

\begin{figure*}[tp]
\hfil\includegraphics[width=\textwidth]{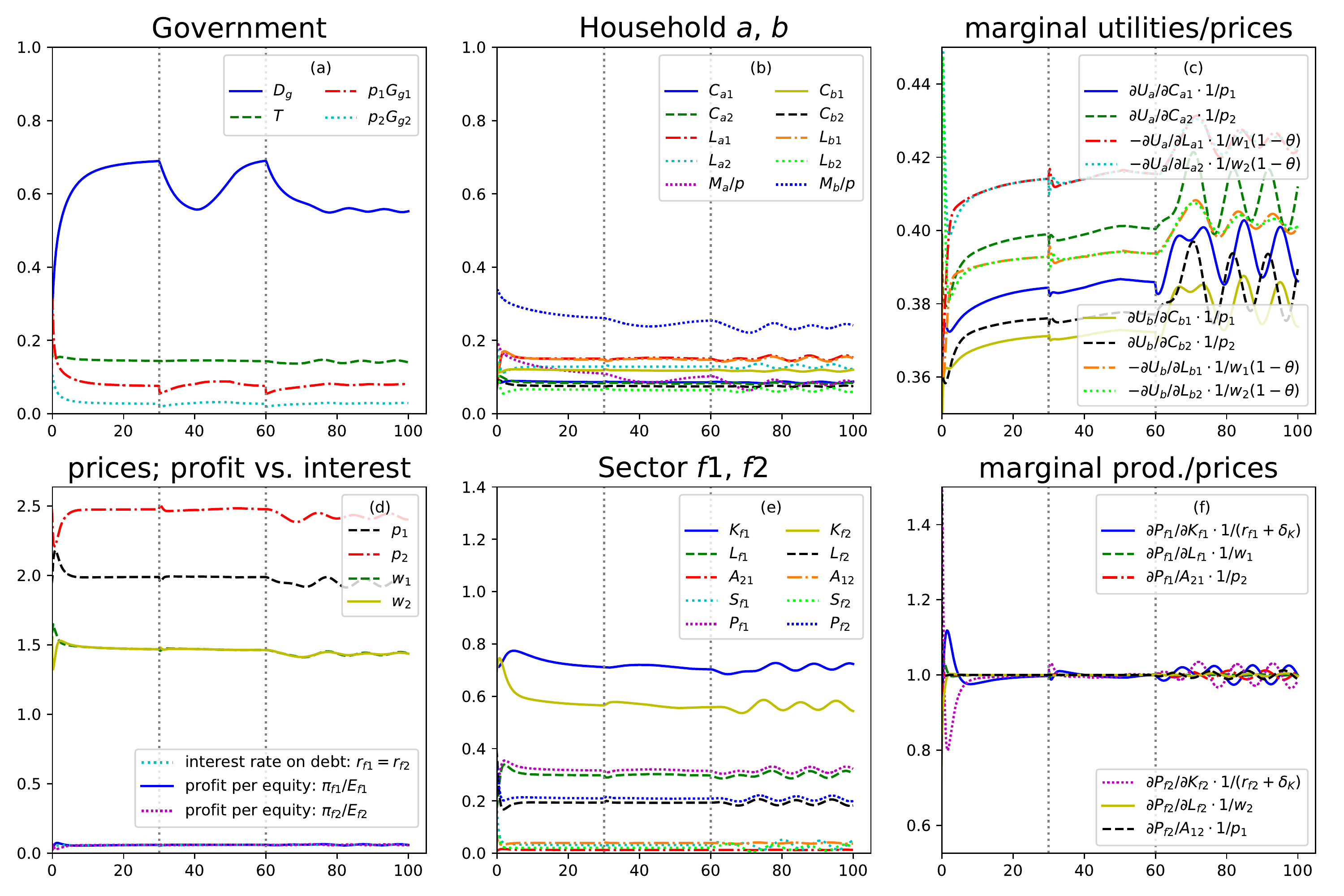}\hfil
\caption{\label{fig:time-evolution-2}Plots (a), (b), (d) and (e) show the time evolution of the variables for different sectors following the specification in Section~\ref{sec_social}. Plot (c) shows that for the two household sectors, the marginal utilities for consumption and leisure, divided by their respective price, do not equalize over time, because we added social interaction.}
\end{figure*}

The results could look as if the framework basically reproduced well-known general equilibrium results for many parameters.
The integration of other variables is straightforward, and Figure~\ref{fig:time-evolution-2} shows a time evolution with social interaction, and changing fiscal and monetary policy.

First, the government is assumed to change to a stronger fiscal conservatism at $t=30$, reflected by a higher aversion $\gamma_D$ to government debt, switching from $0.5$ to $0.6$.
In this part of the timeline, the price adaptation processes are assumed to be sufficiently fast as in Figure~\ref{fig:time-evolution-1}.
In this \emph{neoclassical world}, the lower interest rates push demand for fixed capital, leading to an increase in production.
For $40 < t < 50$, $\gamma_D$ is linearly brought back to $0.5$ such that the second shift in government attitude to debt starts from similar initial conditions.
At $t=60$, the same fast switch to fiscal conservatism occurs, but this time it is assumed that the price adaptation is much slower, all price adaptation speeds are divided by $100$ to make the result visible.
In this \emph{post-Keynesian world}, the reduced government demand reduces sales and therefore depresses investment demand by firms, which is \emph{not} compensated by additional investment demand as the interest rates decrease only slowly.
As a result, the economy goes into recession and starts to oscillate because of the slow price adaptation.
This shows that by adapting the power parameters, the model can continuously shift between the assumptions of different schools of economic thought.

Second, we integrated additional forces to describe social interaction.
Social scientists have emphasized a ``social pressure'' to increase consumption, often framed as ``conspicuous consumption'' to display social status \citep{stiglitz_toward_2008, dutt_happiness_2009, richters_contested_2019}.
In our framework, this social influence can be modeled as a positive influence of households $b$ on consumption decisions by households $a$ (and inversely), by adapting Eqs.~(\ref{model-household-Ca1}--\ref{model-household-Cb2}).
The more household sector $b$ consumes of a certain product, the more household sector $a$ is influenced to increase its consumption.
The power factors $\mu_{baC}$ accounts for the amount of social influence of household sector $b$ on $a$, and $\mu_{abC}$ for the opposite influence.
\begin{align}
\dot C_{a1}\tim &= \mu_{aC1} \cdot \alpha_{C1} (C_{a1}\tim)^{\alpha_{C1}-1}  (C_{a2}\tim)^{\alpha_{C2}} + \mu_{baC} \cdot C_{b1}\tim + \lambda_a\tim p_1\tim - \lambda_{P1}\tim. \label{model-household-Ca1-consp-c} \\
\dot C_{a2}\tim &= \mu_{aC2} \cdot \alpha_{C2} (C_{a1}\tim)^{\alpha_{C1}} (C_{a2}\tim) ^{\alpha_{C2}-1} + \mu_{baC} \cdot C_{b2}\tim + \lambda_a\tim p_2\tim -\lambda_{P2}\tim, \\
\dot C_{b1}\tim &= \mu_{bC1} \cdot \beta_{C1} (C_{b1}\tim)^{\beta_{C1}-1} (C_{b2}\tim)^{\beta_{C2}} + \mu_{abC} \cdot  C_{a1}\tim + \lambda_b\tim p_1\tim -\lambda_{P1}\tim, \\
\dot C_{b2}\tim &= \mu_{bC2} \cdot \beta_{C2} (C_{b1}\tim)^{\beta_{C1}} (C_{b2}\tim)^{\beta_{C2}-1} + \mu_{abC} \cdot  C_{a2}\tim +\lambda_b\tim p_2\tim -\lambda_{P2}\tim.
\end{align}
The result can be seen in Figure~\ref{fig:time-evolution-2}(c), in which the selfish marginal utility divided by the price for consumption goods is much lower than for leisure.
The result of this influence is what \citet{schor_overspent_1999, schor_overworked_1991} describes as ``overspent'' and ``overworked'' consumers.
In a similar way, firms influence on government spending through lobbying or the influence of monopolists on prices can be studied within the framework.

\section{Discussion and conclusions}
\label{sec_conclusion}

This paper presented a dynamic modeling approach in continuous time that extends the analogies between mechanics and economics and depicts the economy from the perspective of economic forces and economic power.
The conceptual model showed how General Constrained Dynamics (GCD) can integrate aspects from behavioral economics, general equilibrium, Keynesian disequilibrium and agent-based models:
It includes some Keynesian features such as slow adaptation of prices and quantities or endogenous money creation.
Similar to agent-based models, the heterogeneous agents have bounded rationality, here modeled as utility improvement by \emph{gradient seeking}.
Nevertheless, in the fixed points of the dynamical system, the first-order conditions of neoclassical general equilibrium solutions are satisfied and the power factors become irrelevant for the production decisions.
The latter can be seen in light of the old debate whether control or economic laws determine market outcomes \citep{von_bohm-bawerk_macht_1914}.

Compared to New Keynesian models with representative or heterogeneous agents, GCD models go back two steps and do without infinite intertemporal optimization and stochastic shocks.
Instead, they allow to study bounded rationality, market inconsistencies far from equilibrium and adaptation processes without the restriction that all utility functions can be aggregated into a social welfare function.\footnote{Note that this question is related to the inverse Lagrangian problem in classical mechanics \citep{douglas_solution_1941, zenkov_inverse_2015}:
For the differential equations provided, generally no Lagrangian function can be determined.}
Different from DSGE models, fast adaptation of quantities and prices does not lead to fast convergence, but can amplify deviations from the equilibrium.
As agents do not react optimally to changing conditions and do not anticipate the reactions of others, frictions have a stabilizing effect.
It remains open whether this result holds if forward looking expectations of firms and the related intertemporal coordination problem are integrated.
If this was the case, political regulation should concentrate on designing market frictions to stabilize markets, instead of eliminating them.

Compared to other disequilibrium models \citep{giraud_household_2019, godley_monetary_2012, chiarella_dynamics_2010, chiarella_quantitative_2006}, the GCD approach allows to explicitly formalize the equilibrating or disequilibrating market forces.
The integration of a Cobb-Douglas production functions enables to study substitution different to many post-Keynesian models that rely on Leontief functions \citep{giraud_household_2019, godley_monetary_2012}, but substitution is modeled as a slow process, not instantaneous adaptation.
Integrating bounded rationality as gradient climbing for households has advantages compared to conventional consumption functions, allowing for a formulation of preferences without assuming that agents are able to optimize.
Still, GCD models have to prove their ability to incorporate the numerous assets and financial entities known from more complex stock-flow consistent models.
The same holds true for a comparison with agent-based models, because the GCD models have not yet been used to describe and analyze high-dimensional systems.
It is doubtful that a system of thousands of differential-algebraic equations arising from thousands of agents can be easily solved.

In the future, the flexibility of the GCD approach allows for various extensions with additional forces.
This includes not only the principal--agent dilemma of dividend policies or credit rationing by banks, but also the influence of monopolists on prices.
Furthermore, political economy issues such as the power relations and influences between politics and firms can be modeled, and the power factor endogenized, for example by making them proportional to wealth as in \citet{makowsky_evolution_2016}.
GCD models may also be applied to further schools of economic thought, for example studying power relation as in Marxist theory or the flows of energy and matter as in integrated assessment models.
By the choice of the parameters that reflect economic power in the sense of the ability to change certain variables and the integration of various social and market forces, the economic and social processes can be modeled in a flexible way.
The presented model contains many ad hoc assumptions which could be refined depending on the application, still the empirical estimation of the power parameters is an unsolved problem.
Further challenges are the integration of only occasionally binding constraints such as a zero lower bound \citep{bohl_efficient_2021}, and the integration of stochastic shocks, bearing in mind that the shocks have to satisfy the economic constraints.
It remains to be shown how forward looking expectations and the intertemporal coordination problem of firms based on bounded rationality could be integrated without any equilibrium assumptions.

In this paper, production and utility functions were chosen such that the dynamics converge to stable equilibria for most parameters.
Economic models with multiple equilibria typically incorporate incomplete markets due to transaction costs or information asymmetries, increasing returns to scale, or market imperfections such as entry costs or external effects \citep{benhabib_chapter_1999}.
They were studied to explain issues such as asset bubbles, collateral shortages, liquidity dry-ups, bank runs, or financial crises \citep{miao_introduction_2016}.
If multiple equilibria exist, a theory that describes the out-of-equilibrium dynamics is required to determine which of the equilibrium states is reached.
A drawback of the GCD approach is that general equilibrium models with multiple markets are tremendously complex in the amount of variables that are simultaneously in equilibrium.
Consequently, providing models able to describe genuine out-of-equilibrium dynamics for all these variables poses a significant challenge.
An intermediate approach could combine equilibrium dynamics with out-of-equilibrium processes where necessary.
As the concept of Lagrangian closure draws on a mathematical similarity to static optimization models, the General Constrained Dynamics framework is a suitable candidate for this task.

\renewcommand{\bibfont}{\normalfont\footnotesize}
\singlespacing
\addcontentsline{toc}{section}{References} 

\printbibliography

\begin{appendices}
\numberwithin{equation}{section}

\section{Initial conditions, power factors and parameters}

\label{sec_initial_conditions}

\paragraph{Initial conditions:}

\noindent Government: expenditures $G_{g1}\timo = 0.05;$ $G_{g2}\timo = 0.02;$ credit $D_g\timo = M_a\timo +M_b\timo - D_{f1}\timo - D_{f2}\timo;$ inflation target $\rho^\top\timo = 0;$

\noindent Household $a$: $L_{a1}\timo = 0.06;$ $L_{a2}\timo = 0.21;$ $C_{a1}\timo = 0.15;$ $C_{a2}\timo = 0.11;$ $M_a\timo = 0.45;$

\noindent Household $b$: $L_{b1}\timo = 0.11;$ $L_{b2}\timo = 0.19;$ $C_{b1}\timo = 0.13;$ $C_{b2}\timo = 0.08;$ $M_b\timo = 0.74;$

\noindent Firms: production inputs $K_{f1}\timo = 0.72;$ $K_{f2}\timo = 0.68;$ $L_{f1} = L_{a1} + L_{b1};$ $L_{f2} = L_{a2} + L_{b2};$ $A_{12}\timo = 0.02;$ $A_{21}\timo = 0.01;$ inventories $S_{f1}\timo = 0.21;$ $S_{f2}\timo = 0.09;$ equity $E_{f1}\timo = 1.07;$ $E_{f2}\timo = 1.71;$ credit $D_{f1} = p_1\timo \left(K_{f1}\timo +S_{f1}\timo \right) - E_{f1}\timo;$ $D_{f2} = p_2\timo \left(K_{f2}\timo +S_{f2}\timo \right) - E_{f2}\timo;$ $E_{bank}\timo = 0$.

\noindent Prices: $w_1\timo = 1.50;$ $w_2\timo = 1.60;$ $p_1\timo = 1.94;$ $p_2\timo = 2.49;$ $r_{f1}\timo  = r_{f2}\timo = r_g\timo = 0.05;$ $r_M\timo = 0.049$.

\paragraph{Power factors:}
\label{sec_power_factors}

\noindent Government: $\mu_{gD} = 2$.

\noindent Household $a$: $\mu_{aL1} = \mu_{aL2} = 2;  \mu_{aC1} = \mu_{aC2} = 2; \mu_{aM} = 2; \mu_{aG1} = 2; \mu_{abC} = 1$.

\noindent Household $b$: $\mu_{bL1} = \mu_{bL2} = 2; \mu_{bC1} = \mu_{bC2} = 2; \mu_{bM} = 2; \mu_{bG2} = 2; \mu_{baC}= 1$.

\noindent Firms: $\mu_{fK1} = 1; \mu_{fK2} = 1; \mu_{fL1} = 2; \mu_{fL2} = 2; \mu_{fA1} = 2; \mu_{fA2} = 2; \mu_{fS1} = 2; \mu_{fS2} = 2$.

\noindent Price development: $\mu_{p1} = 50; \mu_{p2} = 50; \mu_w = 50; \mu_r = 20$.

\paragraph{Parameters:}
\label{sec_parameters}

\noindent Government: utility factors $\gamma_D = 0.5$; $\gamma_r = 0$; tax rate $\theta=0.2$.

\noindent Household $a$: utility factors 
$\alpha_r = 4; \rho_a = 0.06; \alpha_L=0.4; \alpha_{C1}=0.2; \alpha_{C2}=0.25; \alpha_{G1}=0.5$. ownership share $e_a=0.2$.

\noindent Household $b$: utility factors 
$\beta_r = 4; \rho_b = 0.06; \beta_L=0.4; \beta_{C1}=0.25; \beta_{C2}=0.2; \beta_{G2}=0.5$. ownership share $1-e_a$.

\noindent Firms: Cobb-Douglas exponents $\kappa_1=0.25; \kappa_2=0.3; l_1=0.7; l_2 = 0.55$; inventory to sales ratios $s_{f1}^\top = s_{f2}^\top = 0.1$; depreciation $\delta_K=0.05$.

\section{Derivation of firms profits and households income in the stationary state}

\label{appendix-pif1}
Substituting the results from Section \ref{model-stationary} into Eq.~\eqref{model-firm-pif1} yields:
\begin{small}
\begin{align}
\pi_{f1}\eq &= p_1\eq \left( C_{a1}\eq +C_{b1}\eq +G_{g1}\eq +A_{12}\eq \right) - A_{21}\eq p_2\eq -w_1\eq L_{f1}\eq -r_{f1}\eq D_{f1}\eq \\
  &= p_1\eq \left( P_{f1}\eq - \delta_K K_{f1}\eq \right) - A_{21}\eq p_2\eq-w_1\eq L_{f1}\eq - r_{f1}\eq \left[p_1\eq (K_{f1}\eq + s_{f1}^\top P_{f1}\eq)) - E_{f1}\eq\right] \\
  &= p_1\eq P_{f1}\eq \left[1 - (1-r_{f1}\eq s_{f1}^\top) (1-\kappa_1) \right] - (\delta_K + r_{f1}\eq ) p_1\eq K_{f1}\eq - r_{f1}\eq p_1\eq s_{f1}^\top P_{f1}\eq + r_{f1}\eq E_{f1}\eq \\
  &= p_1\eq P_{f1}\eq \left[1 - (1-r_{f1}\eq s_{f1}^\top) (1-\kappa_1) \right] - p_1\eq (1- r_{f1}\eq s_{f1}^\top) \kappa_1 P_{f1}\eq - r_{f1}\eq p_1\eq s_{f1}^\top P_{f1}\eq + r_{f1}\eq E_{f1}\eq \\
  &= p_1\eq P_{f1}\eq \left[1 - (1-r_{f1}\eq s_{f1}^\top) (1-\kappa_1) - (1- r_{f1}\eq s_{f1}^\top) \kappa_1 - r_{f1}\eq s_{f1}^\top \right] + r_{f1}\eq E_{f1}\eq \\
  &= r_{f1}\eq E_{f1}\eq.
\end{align}
\end{small}
Total income distributed to household $a$ and $b$ from production sector $f1$ (interest via the banks) is:
\begin{small}
\begin{align}
& \pi_{f1}\eq + r_{f1}\eq D_{f1}\eq + w_1\eq L_{f1}\eq \\
&\quad = r_{f1}\eq p_1\eq (K_{f1}\eq + s_{f1}^\top P_{f1}\eq) + w_1\eq L_{f1}\eq \\
&\quad = p_1\eq (1-r_{f1}\eq s_{f1}^\top) \kappa_1 P_{f1}\eq - \delta_K p_1\eq K_{f1}\eq + p_1\eq r_{f1}\eq s_{f1}^\top P_{f1}\eq  + p_1\eq (1-r_{f1}\eq s_{f1}^\top) l_1 P_{f1}\eq \\
&\quad = p_1\eq P_{f1}\eq - (1-\kappa_1 - l_1) p_1\eq P_{f1}\eq(1-r_{f1}\eq s_{f1}^\top) - \delta_K p_1\eq K_{f1}\eq \\
&\quad = p_1\eq P_{f1}\eq - p_2\eq A_{21}\eq - \delta_K p_1\eq K_{f1}\eq.
\end{align}
\end{small}

\enlargethispage{\baselineskip}

\end{appendices}

\end{document}